\documentclass[a4paper,12pt]{article}
\usepackage{jheppub}
\usepackage{mathtools}
\usepackage[normalem]{ulem}
\usepackage{orcidlink}
\usepackage{slashed}

\usepackage[compat=1.1.0]{tikz-feynhand}
\tikzset{graviton/.style={decorate, decoration={snake, amplitude=.4mm, segment length=1.5mm, pre length=.5mm, post length=.5mm}, double}}
\tikzset{inflaton/.style={thick, black, dashed}}
\tikzset{reheaton/.style={orange, dashed}}
\tikzset{fermion/.style={thick, black, postaction={decorate},decoration={markings,mark=at position 0.6 with {\arrow{stealth}}}   }}

\newcommand*\circled[1]{\tikz[baseline=(char.base)]{
		\node[shape=circle,draw,inner sep=0.5pt] (char) {#1};}} 
\newcommand{\tcircled}[1]{\raisebox{.5pt}{\textcircled{\raisebox{-.9pt} {#1}}}}

\newcommand{\Trh}{T_\text{rh}}
\newcommand{\arh}{a_\text{rh}}
\newcommand{\Tmax}{T_\text{max}}
\newcommand{\amax}{a_\text{max}}
\newcommand{\aend}{a_I}

\newcommand{\ogw}{\Omega_\text{GW}}
\newcommand{\Hinf}{H_I}

\newcommand{\Teq}{T_\text{eq}}

\title{
	Full-Spectrum Analysis of Gravitational Wave Production from Inflation to Reheating
}
\author[a]{Xun-Jie Xu\orcidlink{0000-0003-3181-1386}}
\author[b]{, Yong Xu\orcidlink{0000-0002-4582-8747}}
\author[c]{, Qiqin Yin\orcidlink{0009-0005-7933-3055}}
\author[a,d]{, and Junyu Zhu\orcidlink{0009-0009-8345-9928}}
\affiliation[a]{Institute of High Energy Physics, Chinese Academy of Sciences, Beijing 100049, China}
\affiliation[b]{PRISMA$^+$ Cluster of Excellence and Mainz Institute for Theoretical Physics\\
	Johannes Gutenberg University, 55099 Mainz, Germany}
\affiliation[c]{School of Physics, Nanjing University, Nanjing 210093, China}
\affiliation[d]{School of Physical Sciences, University of Chinese Academy of Sciences, Beijing 100049, China}

\emailAdd{xuxj@ihep.ac.cn}
\emailAdd{yonxu@uni-mainz.de}
\emailAdd{211870080@smail.nju.edu.cn}
\emailAdd{zhujunyu@ihep.ac.cn}

\abstract{
	In this work, we systematically study gravitational wave (GW) production during both the inflationary and post-inflationary epochs. While inflationary GWs can be readily derived from tensor  perturbations during 
	inflation, post-inflationary GWs 
	arise from a variety of processes during reheating and require detailed treatment for quantitative analysis. We consider four distinct production channels: $(i)$ pure inflaton annihilation, $(ii)$ graviton bremsstrahlung from inflaton decay, $(iii)$ radiation-catalyzed inflaton-graviton conversion, and $(iv)$ scattering among fully thermalized radiation particles. 
	For each channel, we solve the corresponding Boltzmann equation to obtain the GW spectrum and derive a simple yet accurate analytical expression for it. 
	By employing a consistent treatment of all production channels, our analysis yields for the first time the full spectrum of GWs produced during the inflationary and post-inflationary epochs. 
	We find that, while inflationary GWs dominate at low frequencies, post-inflationary processes generally produce high-frequency GWs with considerably high energy densities that may significantly exceed that of inflationary GWs.}

\preprint{MITP-25-036}

\begin{document}
	\maketitle
	
	\section{Introduction}
	Cosmic inflation offers an elegant framework for addressing key problems in standard cosmology~\cite{Starobinsky:1980te, Guth:1980zm, Linde:1981mu, Albrecht:1982wi}, such as the horizon and flatness problems, while also providing a natural source for generating a stochastic background of gravitational waves (GWs)---see \cite{Guzzetti:2016mkm,Giovannini:2019oii} for reviews. 
	In the simplest realization, a single inflaton field undergoes slow-roll evolution along a nearly flat potential, driving a period of exponential expansion. During this phase, quantum fluctuations of the inflaton field generate metric perturbations, sourcing tensor modes. After inflation, as these modes reenter the horizon, they become dynamical and form a primordial GW background \cite{Starobinsky:1979ty, Allen:1987bk, Sahni:1990tx, Turner:1993vb}.
	The exponential expansion stretches these primordial GWs to long wavelengths and low frequencies, with their amplitude directly linked to the inflationary energy scale. Consequently, detecting primordial GWs could offer critical information of inflation; see Refs.~\cite{Caprini:2018mtu, Giovannini:2019oii} for recent reviews for inflationary primordial GWs.
	
	After inflation ends, the vacuum energy of the inflaton field must be transferred to radiations, ultimately forming a thermal bath of Standard Model (SM) particles. This process, known as reheating \cite{Abbott:1982hn, Dolgov:1989us, Kofman:1994rk, Kofman:1997yn} (see also Refs.~\cite{Allahverdi:2010xz, Amin:2014eta, Lozanov:2019jxc, Barman:2025lvk} for reviews), 
	may also play an important role in the production of GWs.
	During reheating, the inflaton oscillates around the minimum of its potential, producing particles coupled to it, including gravitons. 
	Once produced, these gravitons propagate freely through the early universe and, after undergoing cosmological red-shift, eventually form a stochastic background of GWs in the present universe. This is analogous to how photons from the hot Big Bang eventually gave rise to the cosmic microwave background (CMB).
	
	Several specific mechanisms for graviton production during reheating have been investigated in the literature. These include $(i)$ pair production of gravitons via inflaton annihilation \cite{Ema:2015dka, Ema:2016hlw, Ema:2020ggo, Choi:2024ilx, Gross:2024wkl}, $(ii)$ graviton bremsstrahlung via inflaton decay \cite{Nakayama:2018ptw, Huang:2019lgd,Barman:2023ymn, Barman:2023rpg, Kanemura:2023pnv, Bernal:2023wus, Tokareva:2023mrt, Hu:2024awd, Choi:2024acs, Barman:2024htg, Inui:2024wgj, Jiang:2024akb}, $(iii)$ inflaton scattering with the daughter particles produced from inflaton decay \cite{Xu:2024fjl, Bernal:2025lxp}, and $(iv)$ scattering of thermal species~\cite{Ghiglieri:2015nfa,Ghiglieri:2020mhm,Ringwald:2020ist,Bernal:2024jim}.  After reheating, the universe enters a radiation-dominated era, likely composed primarily of the SM plasma, in which the thermal production of gravitons  has been computed in Refs.~\cite{Ghiglieri:2015nfa,Ghiglieri:2020mhm,Ringwald:2020ist}. 
	
	The primary objective of this study is to provide a systematic and consistent analysis of the GW production via the aforementioned mechanisms through the period from inflation to reheating, offering a comprehensive treatment that is currently lacking in the literature. 
	We consider a rather generic framework spanning over vacuum-energy-dominated (VD), matter-dominated (MD), and radiation-dominated (RD) epochs, and analyze the dominant GW production channels in each epoch. 
	For each production channel, we derive a simple yet accurate analytical expression for the GW spectrum. Compared to existing calculations in the literature, a large part of our analytical results are new or contain significant improvements. For instance, by solving the Boltzmann equation of the graviton phase space distribution, we are able to significantly improve the analytical description of the GW energy spectra generated by $(ii)$ and  $(iii)$, which in the previous studies~\cite{Barman:2023ymn,Xu:2024fjl} 
	were calculated by neglecting terms that can be potentially important at high frequencies.
	While previous analytical results for $(ii)$ and $(iii)$ are only valid in the low-frequency regime,  our analytical results are accurate over the entire range. 
	Moreover, for $(iv)$, we demonstrate that the production of gravitons from the SM thermal plasma can be well approximated by a toy model with only three parameters, with the major characteristics such as the logarithmic dependence on the graviton energy arising from $t$-channel scattering fully included. This not only justifies our model-agnostic approach but also offers a simplified yet effective method for calculating graviton production in a Beyond-the-Standard-Model (BSM) thermal bath. 
	
	The remainder of this article is organized as follows. In Sec.~\ref{sec:setup}, we introduce the framework  and sketch out GW production mechanisms considered in this framework. 
	In Sec.~\ref{sec:reheating}, we calculate various background quantities (e.g.,~the temperature, the inflaton and radiation energy densities) which will be used in the subsequent GW calculations.  A comprehensive analysis of GW production is presented in Sec.~\ref{sec:GW}. Our main results are discussed in Sec.~\ref{sec:results}. Finally, we conclude with a summary of our findings in Sec.~\ref{sec:conclusion} and delegate various detailed calculations to appendices.

	\section{The framework \label{sec:setup}}
	
	In this work, we consider a quite general framework that contains
	an inflaton field ($\phi$) and generic radiation species ($R$),
	both minimally coupled to gravity via the action
	\begin{align}
		S=\int d^{4}x\sqrt{-g}\left[\frac{M_{P}^{2}}{2}{\cal R}+\frac{1}{2}g^{\mu\nu}\partial_{\mu}\phi\partial_{\nu}\phi-V(\phi)+\mathcal{L}_{R}\right]\,.\label{eq:action}
	\end{align}
	Here, $g$ is the determinant of the metric $g_{\mu\nu}$ tensor;
	$M_{P}\equiv1/\sqrt{8\pi G_{N}}\simeq2.4\times10^{18}~\text{GeV}$
	is the reduced Planck mass; ${\cal R}$ denotes the Ricci scalar;
	$V(\phi)$ is the potential of the inflaton; and $\mathcal{L}_{R}$
	denotes the Lagrangian of particles that constitute the thermal plasma
	of the radiation-dominated universe. It can be the Lagrangian of the
	Standard Model (SM) or beyond, such as grand unified theories (GUTs)
	or supersymmetric theories. In order to let the inflaton eventually
	release its energy into radiation (i.e., to allow for the reheating
	process), $\phi$ should be coupled to at least one of the particles
	in the $R$ sector. We include such coupling terms into $\mathcal{L}_{R}$.
	The model-dependent content of $\mathcal{L}_{R}$ will be discussed
	later.
	
	Eq.~\eqref{eq:action} gives rise to all gravitational interactions
	responsible for GW production in this work. More specifically,  we
	expand the metric $g_{\mu\nu}$ around the Minkowski metric $\eta_{\mu\nu}=(+,-,-,-)$
	as follows~\cite{Choi:1994ax,Donoghue:1994dn}:
	\begin{align}
		g_{\mu\nu}=\eta_{\mu\nu}+\frac{2}{M_{P}}\,h_{\mu\nu}\,.\label{eq:expansion}
	\end{align}
	In linearized classical General Relativity, the second term $\frac{2}{M_{P}}\,h_{\mu\nu}$
	is regarded as the metric perturbation, reflecting its geometric interpretation.
	We consider an energy scale well below the Planck scale, where the
	Einstein-Hilbert term can be treated within the framework of quantum
	effective field theory \cite{Donoghue:1994dn}. In this context, $h_{\mu\nu}$
	can be interpreted as a dynamical field---namely, the graviton---propagating
	in a Minkowski background, similar to other quantum fields. 
	Given $g_{\mu\nu}$, one can further
	determine the contravariant metric $g^{\mu\nu}$, which appears in
	the Lagrangian densities. Substituting the metric expansion into the
	action leads to the effective interaction between the graviton and
	the energy-momentum tensor \cite{Choi:1994ax,Donoghue:1994dn}: 
	\begin{align}
		\sqrt{-g}
		\mathcal{L}\supset\frac{1}{M_{P}}h_{\mu\nu}\sum_{i}T_{i}^{\mu\nu}\,,\label{eq:effective}
	\end{align}
	where $T_{i}^{\mu\nu}$ represents the energy-momentum tensor of a
	given particle species $i$. 
	
	During different phases of the very early universe, different cosmological
	ingredients play the dominant role in Eq.~\eqref{eq:effective}, accounting
	for GW production at different frequency bands with distinct spectral
	shapes. This work aims at a full-spectrum analysis of the GW production
	from inflation, through reheating, and eventually to radiation domination.
	
\begin{figure}
		\centering
		\includegraphics[width=0.8\textwidth]{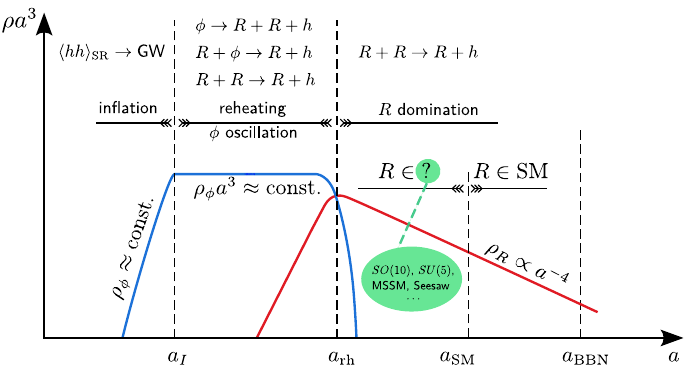}
		\caption{A schematic representation of the three phases from inflation, through
			reheating, and eventually to radiation ($R$) domination. Each phase
			is dominated by a different cosmological ingredient (the scalar potential
			energy, matter, and radiation)  and features GW production with distinct
			spectral shapes.   \label{fig:schematic}}
	\end{figure}

	Figure~\ref{fig:schematic} illustrates the  scope of our framework, where $a$ is the scale factor, and $\rho$ denotes the energy density.
	We start from the inflationary phase, in which the inflaton $\phi$
	undergoes slow-roll (SR) and GWs are produced from the quantum fluctuations
	of $h_{\mu\nu}$ during SR, as indicated by the label ``$\langle hh\rangle_{\text{SR}}\to\text{GW}$''
	in Fig.~\ref{fig:schematic}. Although this is a well-studied subject
	of cosmological inflation, we will revisit the SR production of GWs   
	in this work for completeness and extend the analysis to include horizon crossing during reheating. 
	
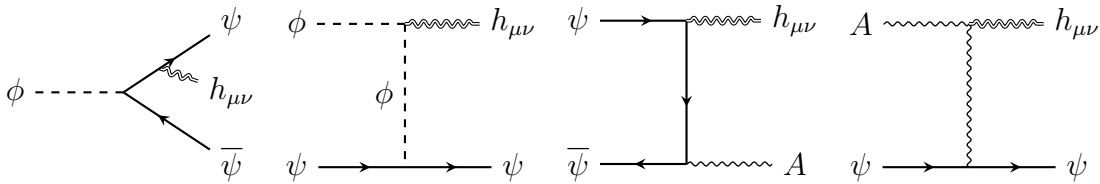
\begin{figure}
		\centering
		\begin{tikzpicture}[scale=0.95]
			\begin{feynhand}
				\vertex (i) at (-1.5, 0) {$\phi$};
				\vertex (M) at (0, 0);
				\vertex (3) at (1.5, 1) {$\psi$};
				\vertex (4) at (1.5, -1) {$\overline{\psi}$};
				\vertex (h) at (1.5, 0) {$h_{\mu\nu}$};
				
				\propag [inflaton] (i) to (M);
				\propag [graviton] (0.5,0.333) to (h);            
				\propag [fermion] (M) to (3);
				\propag [fermion] (4) to (M);
			\end{feynhand}
		\end{tikzpicture}
		\begin{tikzpicture}[scale=0.95]
			\begin{feynhand}
				\vertex (1) at (-1.5, 1) {$\phi$};
				\vertex (2) at (-1.5, -1) {$\psi$};
				\vertex (u) at (0, 1);
				\vertex (d) at (0, -1);
				\vertex (3) at (1.5, 1) {$h_{\mu\nu}$};
				\vertex (4) at (1.5, -1) {$\psi$};
				\propag [inflaton] (1) to (u);
				\propag [fermion] (2) to (d);
				\propag [inflaton] (u) to [edge label'=$\phi$] (d);
				\propag [graviton] (u) to (3);
				\propag [fermion] (d) to (4);
			\end{feynhand}
		\end{tikzpicture}
		\begin{tikzpicture}[scale=0.95]
			\begin{feynhand}
				\vertex (1) at (-1.5, 1) {$\psi$};
				\vertex (2) at (-1.5, -1) {$\overline{\psi}$};
				\vertex (u) at (0, 1);
				\vertex (d) at (0, -1);
				\vertex (3) at (1.5, 1) {$h_{\mu\nu}$};
				\vertex (4) at (1.5, -1) {$A$};
				\propag [fermion] (1) to (u);
				\propag [fermion] (u) to (d);
				\propag [fermion] (d) to (2);
				\propag [photon] (d) to (4);
				\propag [graviton] (u) to (3);
			\end{feynhand}
		\end{tikzpicture}
		\begin{tikzpicture}[scale=0.95]
			\begin{feynhand}
				\vertex (1) at (-1.5, 1) {$A$};
				\vertex (2) at (-1.5, -1) {$\psi$};
				\vertex (u) at (0, 1);
				\vertex (d) at (0, -1);
				\vertex (3) at (1.5, 1) {$h_{\mu\nu}$};
				\vertex (4) at (1.5, -1) {$\psi$};
				\propag [photon] (1) to (u);
				\propag [fermion] (2) to (d);
				\propag [photon] (u) to (d);
				\propag [graviton] (u) to (3);
				\propag [fermion] (d) to (4);
			\end{feynhand}
		\end{tikzpicture}
		
		\caption{Representative Feynman diagrams for graviton production to the order
			of $1/M_{P}$. For a more complete set, see Fig.~\ref{fig:feyn}
			in Appendix~\ref{sec:M2}.\label{fig:Rep-Feyn}}
		
	\end{figure}

	After SR, one of the most likely scenarios is that the inflaton oscillates
	at the bottom of the potential and it is a natural assumption that
	the bottom of the potential is quadratic:
	\begin{align}
		V(\phi)\approx\frac{1}{2}m_{\phi}^{2}\,\phi^{2}\,,\ \ \ \  & (\text{for small }\text{\ensuremath{\phi}})\thinspace.\label{eq:V-quad}
	\end{align}
	This is the case for various inflationary models, including Starobinsky
	inflation \cite{Starobinsky:1980te}, certain classes of $\alpha$-attractor
	T-models \cite{Kallosh:2013hoa,Kallosh:2013maa}, and both small-
	and large-field polynomial inflation scenarios \cite{Drees:2021wgd,Bernal:2021qrl,Drees:2022aea,Drees:2024hok,Bernal:2024ykj}.
	In this phase, the universe is dominated by an oscillating $\phi$
	field, which is physically equivalent to a condensate of $\phi$ particles
	at rest with mass $m_{\phi}$. These $\phi$ particles should eventually
	be converted to radiation in order to accommodate the success of Big
	Bang Nucleosynthesis (BBN). We assume that such conversion is accomplished
	by $\phi$ directly decaying to some generic species $R$. In the
	presence of a decay channel such as $\phi\to R+R$, gravitons ($h$)
	can be produced from bremsstrahlung of the process ($\phi\to R+R+h$)~\cite{Barman:2023ymn}.
	Moreover, as the yield of $R$ increases, $h$ can also be produced
	via $R+\phi\to h+R$ and $R+R\to h+R$. A few of representative Feynman
	diagrams of these processes are illustrated in Fig.~\ref{fig:Rep-Feyn},
	where we consider $R\in\{\psi,\ A\}$ with $\psi$ and $A$ a generic
	fermion and a generic gauge boson. 
	
	After $\phi$ decay, the universe becomes radiation-dominated. In
	this phase, $h$ is produced mainly via scattering of particles in
	the thermal plasma. In the SM plasma, the thermal production of $h$
	has been comprehensively calculated  in Ref.~\cite{Ghiglieri:2015nfa,Ghiglieri:2020mhm}
	(see also \cite{Ringwald:2020ist, Xu:2024cey} for related phenomenological studies).
	In Ref.~\cite{Ghiglieri:2015nfa,Ghiglieri:2020mhm}, scattering between
	any two SM species, including fermions, gauge bosons, and the Higgs
	boson, has been taken into account. Given a variety of potentially
	more fundamental theories beyond the SM at very high energy scales
	(e.g., $SO(10)$, $SU(5)$, MSSM, Seesaw), we adopt a model-agnostic
	approach to study the thermal production of gravitons during the $R$-domination
	phase. In order to simply the analysis, we assume that the plasma
	mainly consists of fermions ($\psi$) and gauge bosons ($A$), with
	their gauge interactions ($\bar{\psi}A_{\mu}\gamma^{\mu}\psi$) responsible
	for maintaining thermal equilibrium and the thermal production of
	$h$. Such a toy model, albeit missing the contribution of scalar
	particles (e.g., the Higgs), is capable to capture the major characteristics
	of graviton production in the SM plasma.  Indeed, as we
	will show, the thermal production rate calculated in this toy model
	agrees well with  the results obtained in the complete SM calculation,
	up to an overall normalization factor that can be accounted for by
	the multiplicity of $\psi$ and $A$. For more unified gauge theories
	such as $SO(10)$ with only one universal gauge coupling for all fermions,
	we expect that the toy model may offer a good and simple approximation
	to the full calculation. 
	
	\section{The post-inflationary background evolution \label{sec:reheating}}
	
	In this section, we briefly discuss the post-inflationary evolution
	of the universe and revisit formulae of background quantities including
	the energy densities of inflaton ($\rho_{\phi}$) and radiation ($\rho_{R}$),
	the temperature $T$, and other relevant variables.

	\subsection{From $a_{I}$ to $a_{\rm rh}$  \label{sub:before-rh}}
	In many inflationary models, it is quite common that after the slow-roll
	phase the inflaton field rolls toward the minimum of its potential
	and undergoes oscillations. The oscillating $\phi$ field with the
	quadratic potential in Eq.~\eqref{eq:V-quad} can be treated as a
	condensate of $\phi$ particles at rest.  Hence its energy density, defined as
	\begin{align}  
		\rho_\phi (\phi) \equiv \frac{\dot{\phi}^2}{2} + V(\phi),  \label{eq:rho-phi-def}
	\end{align}  
	scales as $a^{-3}$ and behaves like matter. Given a certain decay rate of $\phi$
	to radiation, the evolution of $\rho_{\phi}$ and $\rho_{R}$ is governed
	by the following Boltzmann equations: 
	\begin{align}
		& \frac{d\rho_{\phi}}{dt}+3H\rho_{\phi}=-\Gamma_{\phi}\,\rho_{\phi}\,,\label{eq:rhophi}\\
		& \frac{d\rho_{R}}{dt}+4H\rho_{R}=+\Gamma_{\phi}\,\rho_{\phi}\,,\label{eq:rhoR}
	\end{align}
	where $\Gamma_{\phi}$ denotes the energy conversion rate of $\rho_{\phi}$ to $\rho_R$.
	If the thermal effect can be neglected, $\Gamma_{\phi}$ is identical to the decay rate of $\phi$ in vacuum and depends only on the relevant Lagrangian parameters. If the thermal effect is significant, $\Gamma_{\phi}$ can also be temperature-dependent---see e.g.~Eq.~(5) in Ref.~\cite{Kolb:2003ke}.
		Here we only view $\Gamma_{\phi}$ as an effective parameter to quantify the energy conversion rate of $\rho_{\phi}$ to $\rho_R$. More details regarding $\Gamma_{\phi}$ will be discussed in the next section.
	The Hubble parameter is defined as $H\equiv\dot{a}/a$, which is determined by the Friedmann equation:
	\begin{align}
		H^{2}=\frac{\rho_{\phi}+\rho_{R}}{3M_{P}^{2}}\,.\label{eq:Hubble1}
	\end{align}

	When the universe is dominated by the condensate of $\phi$ particles
	(i.e., dominated by matter), the Hubble parameter scales as
	\begin{equation}
		H\approx H_{I}\left(\frac{a_{I}}{a}\right)^{3/2}\thinspace,\label{eq:H-a}
	\end{equation}
	where the subscript ``$I$'' denotes the moment at the end of inflation.
	When using Eq.~\eqref{eq:H-a}, the underlying assumption is that
	\begin{equation}
		\Gamma_{\phi}\ll H_{I}\thinspace,\label{eq:slow-decay}
	\end{equation}
	such that there is a significant matter-dominated period. We refer
	to this as the slow-decaying regime and will use it extensively in
	our analytical calculations below.
	
	Note that $H_{I}$ can be determined by inflationary observables:
	\begin{align}
		\frac{H_{I}}{M_{P}}=1.0\times10^{-5}\left(\frac{\mathcal{P}_{\mathcal{R}}}{2.1\times10^{-9}}\right)^{1/2}\left(\frac{r}{0.01}\right)^{1/2},\label{eq:Hinf}
	\end{align}
	where $r$ is the tensor-to-scalar ratio and $\mathcal{P}_{\mathcal{R}}$
	represents the amplitude of the scalar power spectrum. In this work,
	we remain agnostic about the specific inflationary model and use experimental
	inputs to set $H_{I}$. Adopting the central value $\mathcal{P}_{\mathcal{R}}=2.1\times10^{-9}$
	from Planck 2018~\cite{Planck:2018vyg} and impose the latest constraint
	$r<0.035$ from BICEP/Keck 2018~\cite{BICEP:2021xfz}, we obtain
	\begin{equation}
		H_{I}<1.9\times10^{-5}M_{P}\thinspace.\label{eq:H_I_val}
	\end{equation}
	
	Substituting $H=\dot{a}/a$ into Eq.~\eqref{eq:H-a}, one can solve
	it  as a differential equation of $a(t)$ and obtain
	\begin{align}
		t & \approx\frac{2}{3H_{I}}\left[\left(\frac{a}{a_{I}}\right)^{3/2}-1\right]\thinspace,\label{eq:t-a}\\
		a & \approx a_{I}\left(1+\frac{3}{2}H_{I}t\right)^{2/3}\thinspace.\label{eq:a-t}
	\end{align}
	Here we have set $t=0$ at the moment ``$I$''. 
	
	Using $\frac{d\rho_{\phi}}{dt}+3H\rho_{\phi}=\frac{d\left(\rho_{\phi}a^{3}\right)}{a^{3}dt}$,
	we can write Eq.~(\ref{eq:rhophi}) as $\dot{Y}=-\Gamma_{\phi}\,Y$
	with $Y\equiv\rho_{\phi}a^{3}$, which implies
	\begin{align}
		\rho_{\phi}(a) & =\rho_{\phi}(a_{I})\left(\frac{a_{I}}{a}\right)^{3}e^{-\Gamma_{\phi}t}\label{eq:rhophi_sol}\\
		& \approx3H_{I}^{2}M_{P}^{2}\left(\frac{a_{I}}{a}\right)^{3}\,,\ \ (\text{for }t\lesssim\Gamma_{\phi}^{-1})\thinspace.\label{eq:rhophi_approx}
	\end{align}
	Despite that the $a$-$t$ relation in Eq.~\eqref{eq:t-a} is approximate,
	Eq.~\eqref{eq:rhophi_sol} is exact. Its physical meaning is clear:
	the energy of $\phi$ in a comoving volume ($a^{3}$) decreases exponentially
	because the number of $\phi$ particles in this volume decays exponentially.
	In Eq.~\eqref{eq:rhophi_approx}, we have approximated $e^{-\Gamma_{\phi}t}\approx1$.
	This is valid for $t\lesssim t_{{\rm rh}}$ where $t_{{\rm rh}}$
	is defined as
	\begin{equation}
		t_{{\rm rh}}\equiv\Gamma_{\phi}^{-1}\thinspace.\label{eq:t_rh}
	\end{equation}
	Correspondingly, the scale factor at this point is
	\begin{equation}
		a_{{\rm rh}}\approx a_{I}\left(1+\frac{3}{2}H_{I}t_{{\rm rh}}\right)^{2/3}\approx a_{I}\left(\frac{3H_{I}}{2\Gamma_{\phi}}\right)^{\frac{2}{3}}\thinspace,\label{eq:a_rh}
	\end{equation}
	where in the second step, we have assumed $\frac{3H_{I}}{2\Gamma_{\phi}}\gg1$.
	
	Using Eq.~(\ref{eq:rhophi_approx}), the solution to Eq.~(\ref{eq:rhoR})
	for the radiation energy density is approximately given by 
	\begin{align}
		\rho_{R}(a)\approx\frac{6}{5}M_{P}^{2}\Gamma_{\phi}H_{I}\left(\frac{a_{I}}{a}\right)^{3/2}\left[1-\left(\frac{a_{I}}{a}\right)^{5/2}\right], & \ \ \ (\text{for }t\lesssim t_{{\rm rh}})\thinspace.\label{eq:rho_R_sol}
	\end{align}
	We assume that the radiation,  which is likely to contain
	multiple species as already discussed in the previous section, always
	maintains thermal equilibrium among the multiple species. Therefore,
	it has a well-defined temperature $T$, determined by 
	\begin{align}
		\rho_{R}(T)\equiv\frac{g_{\star}\,\pi^{2}}{30}T^{4}\thinspace,\label{eq:rho-T}
	\end{align}
	where $g_{\star}$ denotes the number of relativistic degrees of freedom
	in the thermal bath. 
	
	Here we shall clarify potential ambiguities of the reheating temperature
	$T_{{\rm rh}}$. If we define it as the temperature of the thermal
	bath at $t=t_{{\rm rh}}$, then by substituting Eq.~\eqref{eq:a_rh}
	into Eq.~\eqref{eq:rho_R_sol} and using Eq.~\eqref{eq:rho-T}, we
	get
	\begin{align}
		T_{{\rm rh}}\approx\frac{2}{\sqrt{\pi}}\left(\frac{3}{2g_{\star}}\right)^{1/4}\sqrt{M_{P}\Gamma_{\phi}}\,.\label{eq:T_rh-2}
	\end{align}
	Alternatively, one can also extrapolate the radiation domination
	backwards until $\rho_{R}$ reaches $3M_{P}^{2}H_{{\rm rh}}^{2}$
	with  $H_{{\rm rh}}\equiv H(t_{{\rm rh}})\approx\frac{2}{3t_{{\rm rh}}}$,
	and define $T_{{\rm rh}}$ as the temperature at this point. This
	gives  
	\begin{align}
		T_{{\rm rh}}\approx\sqrt{\frac{2}{\pi}}\left(\frac{10}{g_{\star}}\right)^{1/4}\sqrt{M_{P}\Gamma_{\phi}}\,.\label{eq:T_rh}
	\end{align}
	The two reheating temperatures only differ by a factor of $0.88$,
	implying that power-law extrapolation from either side is a good approximation.
	In  this work, we choose Eq.~\eqref{eq:T_rh} as
	the definition of  $T_{{\rm rh}}$. 
	
	According to Eq.~\eqref{eq:rho_R_sol}, the radiation energy density
	$\rho_{R}(a)$ reaches its maximum at $a=\amax=(8/3)^{2/5}\aend$,
	corresponding to a maximum temperature 
	\begin{align}
		\Tmax^{4}=\frac{60}{\pi^{2}g_{\star}}\left(\frac{3}{8}\right)^{8/5}M_{P}^{2}\Gamma_{\phi}\Hinf\,.\label{eq:Tmax}
	\end{align}
	Comparing it to $\Trh$ in Eq.~\eqref{eq:T_rh}, we obtain
	\begin{align}
		\frac{\Tmax}{\Trh}\approx0.75\left(\frac{H_{I}}{\Gamma_{\phi}}\right)^{1/4}.\label{eq:Tmax1}
	\end{align}
	Since we are mainly concerned with the slow-decaying regime {[}see
	Eq.~\eqref{eq:slow-decay}{]}, Eq.~\eqref{eq:Tmax1} implies that
	$\Tmax$ in general exceeds $\Trh$~\cite{Giudice:2000ex}. 

	
	\subsection{After $a_{{\rm rh}}$ \label{sub:after-rh}}
	
	After inflatons decay to radiations, the universe becomes radiation-dominated,
	corresponding to the epoch of $a\gtrsim a_{{\rm rh}}$. In this epoch,
	inflatons due to their exponentially suppressed abundance, play a
	negligible role in the production of GWs or gravitons. Instead, radiation-radiation
	scattering becomes the dominant production channel, which has been
	investigated as the thermal production of gravitons in the literature~\cite{Ghiglieri:2015nfa,Ghiglieri:2020mhm,Ringwald:2020ist,Klose:2022knn,Ringwald:2022xif,Klose:2022rxh,Ghiglieri:2022rfp,Muia:2023wru,Drewes:2023oxg,Bernal:2024jim}.
	The thermal production crucially relies on the temperature, with the
	production rate decreasing rapidly as the universe expands and cools
	down. Consequently, one can assume that it ceases producing gravitons
	effectively after the universe expands by a significantly large factor. 
	
	Nevertheless, the subsequent evolution still has an important effect
	on the GW spectrum caused by $g_{\star}$ which decreases from ${\cal O}(100)$
	or higher in the early universe to a value of a few in the present
	universe. The decreasing $g_{\star}$ causes the temperature of the
	thermal bath to be higher than it would be if $g_{\star}$ were a
	constant, thereby increasing the ratio of the mean GW frequancy to
	the CMB frequency.  To take this into account, we use entropy conservation
	to  obtain the relation of $T$ with $a$:
	\begin{equation}
		T=T_{0}\frac{a_{0}}{a}\left(\frac{g_{\star s,0}}{g_{\star s}}\right)^{1/3},\label{eq:T-a}
	\end{equation}
	where $g_{\star s}$ denotes the number effective degrees of freedom
	in entropy, and the subscript ``$0$'' denotes values in the present
	universe. Note that Eq.~\eqref{eq:T-a} remains valid after neutrino
	decoupling, provided that $T$ is interpreted as the temperature of
	photons, not neutrinos. It is also valid in the matter-dominated era
	with $a>a_{{\rm eq}}$ where $a_{{\rm eq}}\approx1/3400$ is the scale
	factor at matter-radiation equality, since matter and dark energy
	do not contribute to the total entropy. 
	
	In this work, we take the following present-day values of relevant
	quantities:
	\begin{equation}
		a_{0}=1\thinspace,\ \ T_{0}=2.73\ {\rm K}\thinspace,\ \ \ \ g_{\star s,0}\approx2+\frac{7}{8}\times\frac{4}{11}\times6\approx3.9\thinspace,\label{eq:today}
	\end{equation}
	\begin{equation}
		H_{0}=100h\ \text{km}\text{s}^{-1}\text{Mpc}^{-1}\ \ \text{with}\ \ h\approx0.67\thinspace.\label{eq:today-H}
	\end{equation}
	The critical energy density is defined as $\rho_{c}=3H_{0}^{2}M_{P}^{2}\simeq1.05\times10^{-5}~h^{2}\ \text{GeV/cm}^{3}$~\cite{ParticleDataGroup:2024cfk}.

	\section{Gravitational wave production}\label{sec:GW}
	With the background evolution determined, we are ready to calculate the production of GWs in our framework covering both the inflationary and post-inflationary epochs.
	
	\subsection{Inflationary GW}
	
	
	During inflation, GWs can be generated by metric perturbations ($h_{\mu\nu}$)
	sourced by quantum fluctuations of the inflaton, corresponding to
	the well-known primordial tensor power spectrum, $\mathcal{P}_{t}$.
	The calculation of the tensor power spectrum in the slow-roll inflationary
	paradigm has been well established (see, e.g., \cite{Caprini:2018mtu,Saikawa:2018rcs})
	and the result is typically formulated as 
	\begin{align}
		\mathcal{P}_{t}=r\,\mathcal{P}_{\mathcal{R}}(k_{\star})\left(\frac{k}{k_{\star}}\right)^{n_{t}}\,,
	\end{align}
	where $r$ denotes the tensor-to-scalar ratio, $n_{t}$ is the tensor
	spectral index, and $\mathcal{P}_{\mathcal{R}}(k_{\star})$ is the
	amplitude of the scalar power spectrum at the pivot scale $k_{\star}$.
	Using the Planck measurement \cite{Planck:2018jri}, we take $k_{\star}=0.05~\text{Mpc}^{-1}$
	and $\mathcal{P}_{\mathcal{R}}(k_{\star})\simeq2.1\times10^{-9}$
	in this work.  In single-field slow-roll inflation, the spectral
	index satisfies $n_{t}\simeq-r/8$. Given the current constraint $r<0.036$
	from BICEP/Keck~\cite{BICEP:2021xfz}, the dependence of $\mathcal{P}_{t}$
	on $k$ is rather weak and therefore is neglected in our work, as
	we aim to minimize the model dependence of our full-spectrum analysis.
	
	The corresponding GW spectrum at the present time is related to the
	primordial tensor power spectrum $\mathcal{P}_{t}$ by a gravitational
	transfer function---see e.g.~\cite{Boyle:2005se,Saikawa:2018rcs}.
	The transfer function encapsulates the expansion history from the
	horizon crossing of a given mode $k$ at $a=a_{{\rm hc}}$ to a later
	time $a>a_{{\rm hc}}$, where $a_{{\rm hc}}$ is the scale factor
	at horizon crossing, determined by 
	\begin{equation}
		a_{{\rm hc}}H(a_{{\rm hc}})=k\thinspace.\label{eq:ah-k}
	\end{equation}
	Different $k$ modes reenter the horizon at different epochs, either
	during radiation domination or reheating. For both epochs, it is straightforward
	to solve Eq.~\eqref{eq:ah-k} and obtain (see Appendix.~\ref{sec:hc}
	for details):
	\begin{equation}
		a_{{\rm hc}}=\begin{cases}
			\sqrt{\frac{\pi^{2}g_{\star,{\rm hc}}}{90}}\left(\frac{g_{\star s,0}}{g_{\star s,{\rm hc}}}\right)^{2/3}\frac{T_{0}^{2}}{kM_{P}} & \ \ \ (k_{{\rm eq}}<k<k_{\text{rh}})\\[2mm]
			\frac{\pi^{2}g_{\star,{\rm rh}}}{90}\left(\frac{g_{\star s,0}}{g_{\star s,{\rm rh}}}\right)\frac{T_{0}^{3}T_{{\rm rh}}}{k^{2}M_{P}^{2}} & \ \ \ (k_{\text{rh}}<k<k_{I})
		\end{cases},\label{eq:ahc}
	\end{equation}
	where the subscripts ``hc'', ``rh'',
	and ``$0$'' in $g_{\star}$ or $g_{\star s}$ indicate that one
	should evaluate $g_{\star}$ or $g_{\star s}$ at $a=a_{{\rm hc}}$,
	$a_{{\rm rh}}$, and $a_{0}$, respectively. 
	The first and second cases in Eq.~\eqref{eq:ahc}
	correspond to $k$ reentering the horizon during radiation domination
	and reheating, respectively. Here $k_{\text{rh}}$, $k_{\text{eq}}$,
	and $k_{I}$ are defined as $k_{X}\equiv a_{X}H(a_{X})$ with $X\in\{\text{rh},\ \text{eq},\ I\}$.
	By equating the two cases with each other, we can obtain $k_{\text{rh}}$
	and the corresponding frequency $f_{\text{rh}}=k_{\text{rh}}/(2\pi)$,
	which reads 
	\begin{equation}
		f_{\text{rh}}=\frac{1}{6}\sqrt{\frac{g_{\star,{\rm rh}}}{10}}\left(\frac{g_{\star s,0}}{g_{\star s,{\rm rh}}}\right)^{\frac{1}{3}}\frac{T_{0}T_{{\rm rh}}}{M_{P}}\simeq3\times10^{4}~\left(\frac{T_{{\rm rh}}}{10^{12}~\text{GeV}}\right)~\text{Hz}\thinspace.\label{eq:frh}
	\end{equation}
	Similarly, for $k_{I}=a_{I}H_{I}$, we find
	\begin{align}
		f_{I} & \equiv\frac{a_{I}H_{I}}{2\,\pi}=\frac{T_{0}}{2}\left(\frac{g_{\star,{\rm rh}}}{90\pi}\thinspace\frac{g_{\star s,0}}{g_{\star s,{\rm rh}}}\thinspace\frac{H_{I}T_{{\rm rh}}}{M_{P}^{2}}\right)^{1/3}\,\nonumber \\
		& \simeq7\times10^{6}\,\left(\frac{T_{{\rm rh}}}{10^{12}~\text{GeV}}\right)^{1/3}\left(\frac{H_{I}}{10^{13}~\text{GeV}}\right)^{1/3}\text{Hz}\,.\label{eq:fI}
	\end{align}
	We note that GWs with frequencies $f>f_{I}$, corresponding to extremely
	short wavelengths, have never exited the horizon. However, this does
	not imply that such modes cannot be excited. As we will discuss in
	the next section, GWs with $f>f_{I}$ can be generated through inflaton-inflaton
	annihilation, $\phi\,\phi\to\,h\,h$.  This shall be understood
	as the high frequency tail of the primordial gravitational waves \cite{Ema:2020ggo,Pi:2024kpw,Giovannini:2025obx}.
	
	By combining the primordial tensor power spectrum and the transfer
	function, one can obtain the present-day GW spectrum sourced by quantum
	fluctuations of the inflaton~\cite{Maggiore:1999vm,Watanabe:2006qe,Saikawa:2018rcs,Caprini:2018mtu}:
	\begin{align}
		\Omega_{{\rm GW}}(k)=\frac{1}{12}\left(\frac{k}{a_{0}H_{0}}\right)^{2}\mathcal{P}_{t}\,\mathcal{T}(a_{0},k)\,,\label{eq:ogw-k}
	\end{align}
	where 
	$\mathcal{T}(a,k)$ is the transfer function accounting for the
	evolution of tensor modes at later times. In the WKB approximation,
	it is given by~\cite{Boyle:2005se,Saikawa:2018rcs}:
	\begin{align}
		\mathcal{T}(a,k)=\frac{1}{2}\left(\frac{a_{{\rm hc}}}{a}\right)^{2},\label{eq:T-fun}
	\end{align}
	where $a_{{\rm hc}}$ depends on $k$, with the explicit form given
	in Eq.~\eqref{eq:ahc}. 
	
	Substituting Eq.~\eqref{eq:T-fun} into Eq.~(\ref{eq:ogw-k}) and
	using Eq.~\eqref{eq:ahc}, we get
	\begin{equation}
		\Omega_{{\rm GW}}(f)=\Omega_{\gamma}^{0}\,\frac{\mathcal{P}_{t}}{24}\,\frac{g_{\star,{\rm hc}}}{g_{\gamma}}\times\begin{cases}
			\left(\frac{g_{\star s,0}}{g_{\star s,{\rm hc}}}\right)^{\frac{4}{3}} & \ \ \ (f_{{\rm eq}}<f<f_{\text{rh}})\\[2mm]
			\frac{g_{\star,{\rm rh}}}{g_{\star,{\rm hc}}}\left(\frac{g_{\star s,0}}{g_{\star s,{\rm rh}}}\right)^{\frac{4}{3}}\left(\frac{f}{f_{{\rm rh}}}\right)^{-2} & \ \ \ (f_{\text{rh}}<f<f_{I})
		\end{cases}.\label{eq:ogw-1}
	\end{equation}
	Here, $\Omega_{\gamma}^{0}\equiv\rho_{\gamma}^{0}/\rho_{c}=2.47\times10^{-5}\,h^{-2}$~\cite{Planck:2018vyg}
	is the present-day photon energy density fraction, and $g_{\gamma}=2$
	accounts for the two degrees of freedom of the photon.    
	
	\subsection{Post-inflationary GW}

	After inflation, the universe is filled with cold $\phi$ particles
	together with subsequently generated thermal radiations. As previously
	illustrated in Fig.~\ref{fig:Rep-Feyn}, gravitons can be produced
	via various particle decay or scattering processes in the post-inflationary
	epoch. These gravitons would constitute a high-frequency GW spectrum,
	with the typical frequency significantly higher than $f_{I}$ in Eq.~\eqref{eq:fI}. 
	
	Since this part is produced via particle decay or scattering processes,
	we adopt the Boltzmann equation to calculate the spectrum. Denoting
	the phase-space distribution function of gravitons by $f_{h}(t,\ k)$
	where $k$ is the graviton momentum, the Boltzmann equation for $f_{h}$
	reads:
	\begin{align}
		\left[\frac{\partial}{\partial t}-H(a)\,k\,\frac{\partial}{\partial k}\right]f_{h}(t,\ k)=\mathcal{C}_{h}(a,\ k)\,,\label{eq:Boltzmann_fh}
	\end{align}
	where $\mathcal{C}_{h}$ is the collision term for graviton production.
	In the freeze-in regime\footnote{Due to the weakness of graviton interactions, the yield of gravitons
		produced via particle processes considered in this work can never
		reach equilibrium ($f_{h}\ll1$),  implying that backreactions of
		these processes can be safely neglected. }, one can assume that it is a function of $a$ and $k$, independent
	of $f_{h}$. Under this assumption,  Eq.~(\ref{eq:Boltzmann_fh})
	can be solved by (see, e.g., \cite{Bernal:2025lxp,Wu:2024uxa,Li:2022bpp}
	for derivations)
	\begin{align}
		f_{h}=\int_{a_{I}}^{a}\frac{da'}{a'\,H(a')}\,\mathcal{C}_{h}\left(a',\ \frac{ak}{a'}\right)\thinspace.\label{eq:Boltzmann_fh-int}
	\end{align}
	We note that Eq.~\eqref{eq:Boltzmann_fh-int} takes into account graviton production throughout the reheating phase and the subsequent radiation-dominated phase with $a > a_{\text{rh}}$. 
		For channels involving the inflaton, the contribution after reheating becomes exponentially suppressed.
		For channels that mainly produces gravitons with $\omega\sim T$, the low- and high-frequency parts of the resulting GW spectrum are typically dominated by earlier and later epochs, respectively. 
		This is because $T$ approximately scales as $a^{-3/8}$ during reheating, while the frequency of a graviton once produced scales as $a^{-1}$. Since the latter decreases faster than the former,  later epochs during reheating generally produce gravitons of higher frequencies than earlier epochs.
	
	Given the phase space distribution function $f_{h}$, we define
	the number density $n_{h}$, the energy density $\rho_{h}$, and its
	differential energy distribution $\frac{d\rho_{h}}{dk}$ as follows:
	\begin{align}
		n_{h} & =g_{h}\int\frac{4\pi\,k^{2}\,dk}{(2\pi)^{3}}\,f_{h}\,,\label{eq:n-h-def}\\
		\rho_{h} & =g_{h}\int\frac{4\pi\,k^{3}\,dk}{(2\pi)^{3}}\,f_{h}\,,\label{eq:rGW-int}\\
		\frac{d\rho_{h}}{dk} & \equiv g_{h}\frac{4\pi\,k^{3}}{(2\pi)^{3}}\,f_{h}\,,\label{eq:drho-dk}
	\end{align}
	where $g_{h}=2$ is the number of graviton degrees of freedom.  The
	GW spectrum at present, as a function of the frequency $f=k/(2\pi)$,
	is then given by
	\begin{align}
		\Omega_{{\rm GW}}(f)\equiv\frac{1}{\rho_{c}}\,\frac{d\rho_{h}}{d\ln k}=8\pi^{2}g_{h}\,\frac{f^{4}}{\rho_{c}}\,f_{h}\,,\label{eq:GW_definition}
	\end{align}
	where $f_{h}$ should be evaluated at the present time.

	Finally, we shall briefly introduce the collision term $\mathcal{C}_{h}$
	in the Boltzmann equation. For a general scattering process, $X_{1}+X_{2}+\cdots+X_{n}\to X_{n+1}+X_{n+2}+\cdots+X_{n+m}+h,$
	neglecting quantum statistics factors (e.g., the Pauli blocking factor),
	the collision term for the production of $h$ is given by the phase
	space integral 
	\begin{equation}
		\mathcal{C}_{h}=\frac{g_{n+m}}{2\omega}\int\left(\prod_{i=1}^{n+m}d\Pi_{i}\right)\,f_{1}\cdots f_{n}\,{\cal S}|\overline{\mathcal{M}}|^{2}\,(2\pi)^{4}\delta^{(4)}\left(\sum_{i=1}^{n}p_{i}-\sum_{j=n+1}^{n+m}p_{j}\right),\label{eq:collision-def}
	\end{equation}
	with
	\begin{equation}
		d\Pi_{i}\equiv\frac{d^{3}\mathbf{p}_{i}}{(2\pi)^{3}2E_{i}}\thinspace.\label{eq:dPi}
	\end{equation}
	Here, $\omega$ is the energy of the graviton, $p_{i}$ and $E_{i}$
	denote the momentum and energy of the $i^{\text{th}}$ particle, $f_{i}$
	is its phase-space distribution, $|\overline{\mathcal{M}}|^{2}$ is
	the squared amplitude of this process (with an average taken over
	the spins and polarizations of all initial and final state particles),
	${\cal S}$ is the symmetry factor, and the delta function $(2\pi)^{4}\delta^{(4)}(\dots)$
	ensures energy-momentum conservation. Note that we have factored out
	the multiplicity factors of all particles (except for $h$) from the
	integral in order to make the dependence on these factors explicit
	and define
	\begin{equation}
		g_{n+m}\equiv\prod_{i=1}^{n+m}g_{i}\thinspace,\label{eq:gnm}
	\end{equation}
	where $g_{i}$ is the multiplicity factor of the $i^{\text{th}}$
	particle. The multiplicity factor of the graviton, $g_{h}=2$, is
	not included in Eq.~\eqref{eq:gnm} because $f_{h}$ in Eq.~\eqref{eq:Boltzmann_fh}
	does not include $g_{h}$---see also Eq.~\eqref{eq:rGW-int} in which
	$g_{h}$ is eventually included. 
	
	Below we start our analyses for specific processes, using the above
	prescription to calculate their collision terms and the corresponding
	$f_{h}$ and $\Omega_{{\rm GW}}$. 
	
	\subsubsection{GWs from inflaton annihilation\label{subsec:phi-phi}}
	
	Well before the inflaton starts decaying, the post-inflationary universe
	is exclusively filled with non-relativistic $\phi$ particles. In
	this epoch, gravitons are produced via the inflaton annihilation channel:
	$\phi\,\phi\to\,h\,h$. This production channel has been previously
	studied in Refs.~\cite{Choi:2024ilx,Bernal:2025lxp} and the squared
	matrix element reads (see Appendix~\ref{sec:M2} for the detailed
	calculation):
	\begin{align}
		|\overline{\mathcal{M}}_{\phi\phi\to hh}|^{2}=\frac{2\,m_{\phi}^{4}}{M_{P}^{4}}\times\frac{1}{4}\,,\label{eq:M2-anni}
	\end{align}
	where the factor of $1/4$ arises from taking the average of the polarizations
	of the two gravitons\footnote{Note that in Eq.~(10) of Ref.~\cite{Choi:2024ilx}, the squared
		matrix element is also multiplied by a factor of $1/4$. But that
		one is a symmetry factor, corresponding to our ${\cal S}$ in Eq.~\eqref{eq:collision-def}.
		Our factor of $1/4$, arising from polarization averaging, will eventually
		be canceled out by two $g_{h}$'s when $|\overline{\mathcal{M}}_{\phi\phi\to hh}|^{2}$
		is used to compute the production rate for $\rho_{h}$ or $n_{h}$---see
		Eq.~\eqref{eq:int-C} below, which agrees with Eq.~(11) of Ref.~\cite{Choi:2024ilx}. }. Substituting Eq.~\eqref{eq:M2-anni} into Eq.~(\ref{eq:collision-def})
	and performing the phase space integral (see Appendix~\ref{sec:Calc-collision}
	for details), we obtain
	\begin{align}
		\mathcal{C}_{h} & =\frac{\pi}{32}\,\frac{n_{\phi}^{2}}{M_{P}^{4}}\,\delta(\omega-m_{\phi})\,.\label{eq:Ch-anni}
	\end{align}
	Here $n_{\phi}$ is
	the number density of $\phi$ and we have taken into account the symmetry
	factor as well as double graviton production. The $\delta(\omega-m_{\phi})$
	function indicates that the produced graviton is monochromatic. 
	Due to the monochromatic spectrum, the Boltzmann equations of $\rho_{h}$
	and $n_{h}$ share the same production rate:
	\begin{align}
		\dot{\rho}_{h}+4H\rho_{h} & =\Gamma_{h}\rho_{\phi}\thinspace,\label{eq:rho-Boltz}\\
		\dot{n}_{h}+3Hn_{h} & =\Gamma_{h}n_{\phi}\thinspace,\label{eq:n-Boltz}
	\end{align}
	with
	\begin{equation}
		\Gamma_{h}\equiv\frac{g_{h}}{n_{\phi}}\int\mathcal{C}_{h}\frac{d^{3}k}{(2\pi)^{3}}=\frac{g_{h}n_{\phi}m_{\phi}^{2}}{64\pi M_{P}^{4}}\thinspace.\label{eq:int-C}
	\end{equation}
	
	At the end of reheating, the graviton phase-space distribution obtained
	by calculating the integral in Eq.~(\ref{eq:Boltzmann_fh-int}) is
	given by: 
	\begin{align}
		f_{h}(\arh,k) & \simeq\frac{9\pi}{32}\left(\frac{H_{I}}{m_{\phi}}\right)^{3}\left(\frac{a_{I}}{\arh}\,\frac{m_{\phi}}{k}\right)^{9/2}
	\end{align}
	where we have used $n_{\phi}\simeq n_{\phi I}a_{I}^{3}/a^{3}e^{-\Gamma_{\phi}t}$ 
	during reheating,
	with $n_{\phi I}$ the initial value of $n_{\phi}$. 
	
	Assuming that this channel effectively stops producing $h$ when $a>a_{{\rm rh}}$,
	the present-day distribution $f_{h}(a_{0},k)$ is related to $f_{h}(a_{{\rm rh}},k)$
	by simple red-shifting:
	\begin{equation}
		f_{h}(a_{0},k)=f_{h}(a_{{\rm rh}},ka_{0}/a_{{\rm rh}})\thinspace.\label{eq:f-redshift}
	\end{equation}
	Then using  Eq.~\eqref{eq:GW_definition}, we find 
	\begin{align}\label{eq:ogw_inflaton_inflaton}
		\Omega_{{\rm GW}}h^{2}(f)\simeq9\cdot10^{-22}\left(\frac{m_{\phi}}{10^{13}\thinspace\text{GeV}}\cdot\frac{\Trh}{10^{13}\thinspace\text{GeV}}\right)^{\frac{3}{2}}\left(\frac{\text{GHz}}{f}\right)^{\frac{1}{2}}
		.
	\end{align}
	We remind the reader that the graviton energy at production is the
	same as the inflaton mass, i.e., $\omega(a_{p})=m_{\phi}$, with $a_{p}$
	the scale factor at production. This implies that the graviton energy
	at the end of reheating lies in the range $m_{\phi}a_{I}/a_{{\rm rh}}\leq\omega\leq m_{\phi}$.
	After taking the red-shift from $a_{{\rm rh}}$ to $a_{0}$ into account,
	the GW energy at present should be red-shifted by a factor is $a_{{\rm rh}}/a_{0}$,
	leading to the following frequency band: 
	\begin{equation}
		f_{\text{min}}\leq f\lesssim f_{\text{max}}\,,\label{eq:f_range_2to2}
	\end{equation}
	where 
	\begin{align}
		f_{\text{min}} = \frac{m_{\phi}}{2\pi}\frac{a_{I}}{a_{0}} & \simeq7\cdot10^{6}\,\left(\frac{m_{\phi}}{10^{13}~\text{GeV}}\right)\,\left(\frac{\Trh}{10^{12}~\text{GeV}}\right)^{1/3}\left(\frac{10^{13}~\text{GeV}}{H_{I}}\right)^{2/3}\text{Hz}\,,\label{eq:fmin}\\
		f_{\text{max}} = \frac{m_{\phi}}{2\pi}\frac{a_{{\rm rh}}}{a_{0}}  & \simeq2\cdot10^{11}\,\left(\frac{m_{\phi}}{10^{13}~\text{GeV}}\right)\,\left(\frac{10^{12}~\text{GeV}}{\Trh}\right)\text{Hz}\,.\label{eq:fmax}
	\end{align}
	Note that gravitons with frequencies above $f_{\text{max}}$ can,
	in principle, be produced after reheating; however, their amplitude
	is exponentially suppressed due to the rapid depletion of the inflaton number density. 
	
	\subsubsection{GWs from bremsstrahlung of inflaton decay\label{subsec:brem}}
	
	In our framework, the inflaton $\phi$ is unstable and has a decay
	rate, $\Gamma_{\phi}$, which is essential to reheating. The specific
	form of $\Gamma_{\phi}$ depends on how $\phi$ is coupled to lighter
	species (radiation). Although this is model-dependent, for any given
	decay channel of $\phi$, one can generally expect that a graviton
	can be emitted via bremsstrahlung of the decay channel, and the bremsstrahlung
	decay rate is roughly given by
	\begin{equation}
		\Gamma_{\phi\to h+\cdots}\sim\frac{1}{16\pi^{2}}\frac{m_{\phi}^{2}}{M_{P}^{2}}\Gamma_{\phi\to\cdots}\thinspace\label{eq:brem-estimate}
	\end{equation}
	where ``$\cdots$'' denotes arbitrary final states of the given
	decay channel. The factor of $1/(16\pi^{2})$ is a typical phase-space
	suppression factor of bremsstrahlung, and $m_{\phi}^{2}/M_{P}^{2}$
	accounts for the suppression of a gravitational interaction vertex. 
	
	Eq.~\eqref{eq:brem-estimate} offers a crude estimation of the bremsstrahlung
	decay rate, which may deviate from the exact value by a factor of
	a few, depending on whether $\phi$ decays to scalar/vector bosons
	or fermions~\cite{Barman:2023ymn}. For concreteness, we focus on
	the scenario  that $\phi$ dominantly decays to fermions via a Yukawa
	interaction:
	\begin{equation}
		{\cal L}\supset y\overline{\psi}\psi\phi\thinspace,\label{eq:yukawa}
	\end{equation}
	where $y$ is the Yukawa coupling and $\psi$ is a light fermion with
	a negligible  mass.  We note that the couplings between the inflaton and the daughter fields induce effective mass terms for the latter, proportional to the inflaton field value $\phi$, thereby altering the decay kinematics \cite{Ichikawa:2008ne, Drewes:2019rxn,Garcia:2020wiy}. To properly account for this effect, one shall average over the inflaton oscillations, which yields an effective coupling $y_{\text{eff}}$ the interaction in Eq.~\eqref{eq:yukawa}. In the case of the quadratic inflaton potential considered here, this effect has been shown to be moderate, with $y_{\text{eff}} \simeq y$ \cite{Ichikawa:2008ne, Garcia:2020wiy}.

	Eq.~\eqref{eq:yukawa} gives rise to the dominant decay channel $\phi\to\overline{\psi}\psi$
	with the decay rate
	\begin{equation}
		\Gamma_{\phi}=\frac{y^{2}}{8\pi}m_{\phi}\thinspace.\label{eq:decay-rate-0}
	\end{equation}
	Here, we have assumed an inflaton decay rate in vacuum. It has been shown that backreaction effects on the inflaton decay can arise if the daughter field acquires a thermal mass~\cite{Kolb:2003ke, Yokoyama:2005dv, Drewes:2010pf, Mukaida:2012bz}. Throughout our numerical analysis, we will restrict ourselves to scenarios in which the reheating temperature remains smaller than the inflaton mass. This is typically the case for a small inflaton decay rate (or equivalently, small Yukawa couplings). In such cases, assuming a vacuum inflaton decay rate and neglecting the thermal mass of the daughter field is a good assumption.
	Incorporating it into the Einstein-Hilbert action, Eq.~\eqref{eq:yukawa}
	also implies that gravitons can be produced via the bremsstrahlung decay channel, $\phi\to\overline{\psi}\psi h$,
	for which the  squared matrix element reads (see Appendix~\ref{sec:M2}):
	\begin{align}
		|\overline{\mathcal{M}}_{\phi\to\bar{\psi}\psi h}|^{2}=\frac{1}{8}\times\frac{y^{2}\,m_{\phi}^{2}}{M_{P}^{2}}\left(1-\frac{2\omega}{m_{\phi}}\right)\left[2-\frac{2m_{\phi}}{\omega}+\left(\frac{m_{\phi}}{\omega}\right)^{2}\right]\,.\label{eq:M_Brem}
	\end{align}
	Here the factor of $1/8$ arises from averaging the spins and polarizations
	of the final states. 
	
	Note that, for fixed $\omega$, Eq.~\eqref{eq:M_Brem} is independent
	of the energy of $\psi$. This can be used to greatly simplify the
	calculation of the corresponding collision term (see Appendix~\ref{sec:Calc-collision}
	for details). The resulting collision term is
	\begin{equation}
		\mathcal{C}_{h}=\frac{y^{2}\rho_{\phi}}{64\pi\omega M_{P}^{2}}\left(1-\frac{2\omega}{m_{\phi}}\right)\left[2-\frac{2m_{\phi}}{\omega}+\left(\frac{m_{\phi}}{\omega}\right)^{2}\right]\Theta\left(\frac{m_{\phi}}{2}-\omega\right),\label{eq:Ch-brem-decay}
	\end{equation}
	where $\Theta$ denotes the Heaviside theta function, reflecting the
	kinematical constraint that the graviton produced from a three-body
	decay process must have $\omega<m_{\phi}/2$. 
	
	Solving the Boltzmann equation using Eq.~(\ref{eq:Boltzmann_fh-int}),
	we find
	\begin{align}
		f_{h}(a_{{\rm rh}},k) & \simeq\frac{y^{2}\,H_{I}\,m_{\phi}^{2}}{32\,\pi\,k^{3}}r_{a}^{3/2}\times\begin{cases}
			{\cal J}\left(\frac{k}{m_{\phi}}\right) & \text{for}\ \frac{r_{a}}{2}<\frac{k}{m_{\phi}}<\frac{1}{2}\\
			{\cal J}\left(\frac{k}{m_{\phi}}\right)-r_{a}^{3/2}{\cal J}\left(\frac{k}{r_{a}m_{\phi}}\right) & \text{for}\ \frac{k}{m_{\phi}}<\frac{r_{a}}{2}
		\end{cases}\,,\label{eq:fh-brem-decay}
	\end{align}
	with $r_{a}\equiv a_{I}/a_{{\rm rh}}$ and
	\begin{equation}
		{\cal J}(x)\equiv1-12x+18\sqrt{2}x^{3/2}-18x^{2}+4x^{3}\thinspace.\label{eq:J-x}
	\end{equation}
	The piecewise feature of Eq.~\eqref{eq:fh-brem-decay} originates
	from the $\Theta$ function in Eq.~\eqref{eq:Ch-brem-decay}. The
	${\cal J}(x)$ function monotonically decreases from $1$ to $0$
	for $x\in[0,1/2]$.

	Then using Eq.~\eqref{eq:f-redshift} and Eq.~\eqref{eq:GW_definition},
	we obtain the GW spectrum:
	\begin{align}
		\Omega_{{\rm GW}}h^{2}(f) & \simeq6.2\times10^{-17}
		\cdot\frac{m_{\phi}}{10^{13}~\text{GeV}}
		\cdot\frac{f}{\text{GHz}}\cdot
		g_{\star106}^{-\frac{1}{4}}
		\left(\frac{\Gamma_{\phi}}{10^{-5}M_{P}}\right)^{\frac{1}{2}}
		{\cal J}\left(\frac{f}{f_{\text{max}}}\right),\label{eq:ogw-bremJ}
	\end{align}
	with $g_{\star106}\equiv g_{\star,{\rm rh}}/106.75$, and   $f_{\text{max}}$ corresponding  to the frequency of a graviton produced at $a=a_{{\rm rh}}$ with $\omega=m_{\phi}$ and red-shifted to $a=a_{0}$; see Eq.~\eqref{eq:fmax}.
	In deriving Eq.~\eqref{eq:ogw-bremJ}, we have replaced $y^2$ by $8\pi m_{_\phi}/\Gamma_{\phi}$  and expressed $T_{\rm rh}$ in terms of $\Gamma_{\phi}$ 
	according to Eqs.~\eqref{eq:decay-rate-0} and (\ref{eq:T_rh}). 
	For a crude estimate, one can ignore the factor ${\cal J}(f/f_{\text{max}})$
	in Eq.~\eqref{eq:ogw-bremJ} because it is typically of ${\cal O}(1)$.
	Eq.~\eqref{eq:ogw-bremJ}  is valid only for $f\leq f_{\text{max}} 2$ and
	it vanishes at $f=f_{\text{max}}/2$. However, higher frequency GWs with $f>f_{\text{max}}/2$
	can still be generated via the bremsstrahlung channel, due to a small
	amount of inflatons decaying in the radiation-dominated epoch, but
	this contribution is exponentially suppressed, similar to that discussed
	at the end of Sec.~\ref{subsec:phi-phi}.
	
	\subsubsection{Radiation-catalyzed inflaton-graviton conversion\label{subsec:catalyzed}
	}
	
	In the presence of the decay channel $\phi\to\bar{\psi}\psi h$ discussed
	above, the accompanying process $\phi\psi\to\psi h$ is also possible,
	provided that a significant amount of $\psi$ particles have already
	been produced from $\phi$ decay. In this process, $\psi$ can be
	viewed as a catalyst that facilitates the conversion of $\phi$ to
	$h$, with itself not consumed after the reaction. This radiation-catalyzed
	inflaton-graviton conversion can be very efficient, causing an enhanced
	GW production rate compared to the bremsstrahlung decay under certain
	circumstances, as we will show later.
	
	Note that this process shares essentially the same Feynman diagrams
	as $\phi\to\bar{\psi}\psi h$: pulling the $\bar{\psi}$ leg in the
	diagrams for the latter from final states to initial state yields
	exactly the diagrams for the latter. Hence the squared matrix element
	of $\phi\psi\to\psi h$ can be obtained  from that of $\phi\to\bar{\psi}\psi h$
	using crossing symmetry---see Appendix~\ref{sec:M2}. The result
	is given as follows:
	\begin{align}
		|\overline{\mathcal{M}}_{\phi\psi\to\psi h}|^{2}=\frac{1}{8}\times\frac{y^{2}\,m_{\phi}^{2}}{M_{P}^{2}}\left(\frac{2\omega}{m_{\phi}}-1\right)\left[2-\frac{2m_{\phi}}{\omega}+\left(\frac{m_{\phi}}{\omega}\right)^{2}\right].\label{eq:M_phi_psi}
	\end{align}
	The subsequent calculation of $\mathcal{C}_{h}$ and $f_{h}$ is
	straightforward and similar to the previous one, leading to 
	\begin{align}
		\mathcal{C}_{h} & =\frac{|\overline{\mathcal{M}}_{\phi\psi\rightarrow\psi h}|^{2}n_{\phi}}{8\pi m_{\phi}\omega^{2}}\left(\int_{\omega-m_{\phi}/2}^{\infty}dp_{\psi}f_{\psi}\right)\Theta(\omega-m_{\phi}/2)\label{eq:Ch-catalyze-0}\\
		& \approx\frac{y^{2}\rho_{\phi}}{64\pi\omega M_{P}^{2}}\left(\frac{2\omega}{m_{\phi}}-1\right)\left[2-\frac{2m_{\phi}}{\omega}+\left(\frac{m_{\phi}}{\omega}\right)^{2}\right]\frac{T}{\omega}e^{-\frac{\omega-m_{\phi}/2}{T}}\Theta\left(\omega-\frac{m_{\phi}}{2}\right)\thinspace,\label{eq:Ch-catalyze}
	\end{align}
	where $f_{\psi}(p_{\psi})$ denotes the phase space distribution function
	of $\psi$ and in the last step we have used the Boltzmann approximation:
	$f_{2}\approx e^{-p_{2}/T}$. 
	
	Substituting Eq.~\eqref{eq:Ch-catalyze} into Eq.~\eqref{eq:Boltzmann_fh-int}
	cannot yield an analytically calculable integral. However,  we observe
	that, in practice, one can take $\frac{2\omega}{m_{\phi}}\gg1$ in
	Eq.~\eqref{eq:Ch-catalyze} to obtain a simple yet accurate analytical
	expression for $f_{h}$: 
	\begin{equation}
		f_{h}(a_{{\rm rh}},k)\approx\frac{y^{2}H_{I}\,T_{{\rm rh}}}{2\pi m_{\phi}^{2}}\left(1-\frac{r_{a}m_{\phi}}{2k}\right)^{3}r_{a}^{\frac{3}{2}}\frac{3m_{\phi}}{5k}\left(\frac{T_{\text{rh}}}{k}\right)^{\frac{7}{5}}\Upsilon_{\frac{7}{5}}\left(\frac{k}{T_{\text{rh}}}\right),\label{eq:f-cata}
	\end{equation}
	with
	\begin{equation}
		r_{a}\equiv\frac{a_{I}}{a_{{\rm rh}}}\approx\left(\frac{2\Gamma_{\phi}}{3H_{I}}\right)^{\frac{2}{3}}.\label{eq:ra-def}
	\end{equation}
	In deriving the above $f_{h}$, we have used $\rho_{R}\propto a^{-3/2}$
	{[}see Eq.~\eqref{eq:rho_R_sol}{]}, which implies $T\approx T_{{\rm rh}}(a_{{\rm rh}}/a)^{3/8}$
	during reheating. The $\Upsilon$ function used in Eq.~\eqref{eq:f-cata} [also used in Eq.~\eqref{eq:f-RR} below]
	is defined as follows:
	\begin{equation}
		\Upsilon_{\xi}\left(x\right)\equiv\Gamma\left(\xi,\ x\right)-\Gamma\left(\xi,\ r_{a}^{-\frac{5}{8}}\, x\right),\label{eq:ups}
	\end{equation}
	where $\Gamma$ is the incomplete gamma function.

	\begin{figure}
		\centering
		
		\includegraphics[width=0.6\textwidth]{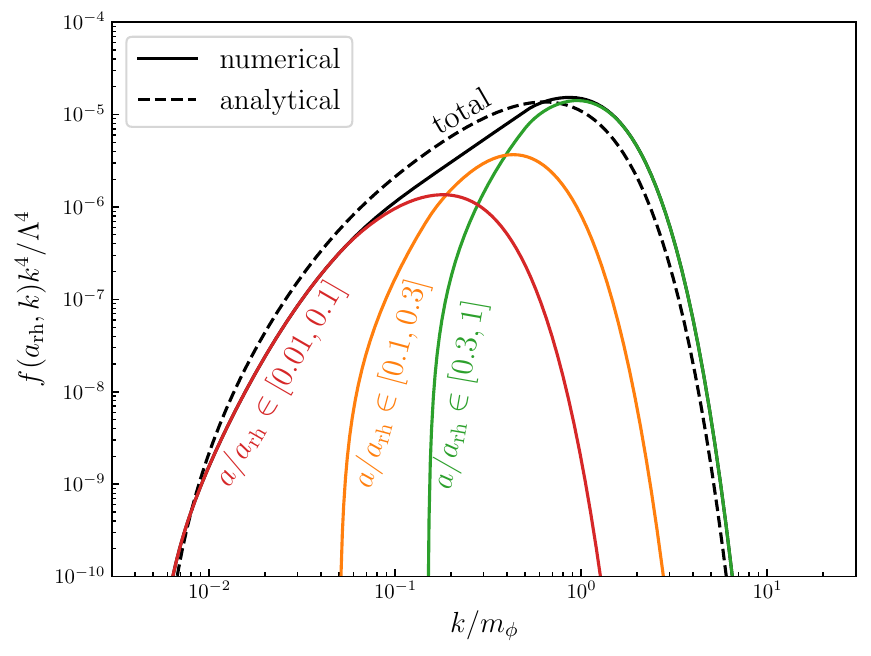}
		
		\caption{Comparison of Eq.~\eqref{eq:f-cata} with the numerical result obtained
			by performing the integration in Eq.~\eqref{eq:Boltzmann_fh-int} with
			${\cal C}_{h}$ in Eq.~\eqref{eq:Ch-catalyze}. Since $\Omega_{{\rm GW}}\propto k^{4}f_{h}$,
			we multiply $f_{h}$ by a dimensionless factor $k^{4}/\Lambda^{4}$
			with $\Lambda^{4}\equiv\frac{y^{2}}{2\pi}H_{I}\,T_{{\rm rh}}m_{\phi}^{2}$.
			In the shown example,
			we have used $r_{a}=10^{-2}$ and $T_{{\rm rh}}/m_{\phi}=1/3$. 
			\label{fig:cata}}
		
	\end{figure}
	
	Since Eq.~\eqref{eq:f-cata} relies on $\frac{2\omega}{m_{\phi}}\gg1$
	while the thermal average of $\omega$ is typically about $3T$, we
	expect that Eq.~\eqref{eq:f-cata} is accurate when 
	\begin{equation}
		T_{{\rm rh}}\gg\frac{1}{6}m_{\phi}\thinspace.\label{eq:xT6}
	\end{equation}

	
	Figure~\ref{fig:cata} shows a comparison of Eq.~\eqref{eq:f-cata} with the numerical result and demonstrates modest accuracy of our analytical calculation. In the shown example, $r_a=10^{-2}$, implying that the universe expands by two orders of magnitude from the end of inflation to the end of reheating. In order to gain a better understanding of when the spectrum receives the dominant contribution, we decompose it into contributions from three different epochs: $a/a_{\rm rh} \in [0.01,0.1]$, $a/a_{\rm rh} \in [0.1,0.3]$, and $a/a_{\rm rh} \in [0.3,1]$. They are presented in Fig.~\ref{fig:cata} by colored lines. From the decomposition, one can see that the spectrum at its peak receives the dominant contribution from the last epoch. Earlier epochs mainly contribute to the low-frequency part of the spectrum due to redshift.

	Using Eq.~\eqref{eq:f-cata}, we obtain the following the GW spectrum
	\begin{equation}
		\Omega_{{\rm GW}}h^{2}(f)\simeq4\times10^{-9}\left(\frac{10^{13}~\text{GeV}}{m_{\phi}}\cdot\frac{\Gamma_{\phi}}{10^{-5}M_{P}}\right)^{2}\left(\frac{f}{10^{9}~\text{Hz}}\right)^{\frac{8}{5}}
		\frac{\Upsilon_{\frac{7}{5}}(f/f_c)}{g_{\star106}^{4/5}}
		\thinspace,\label{eq:ogw-cata}
	\end{equation}
	where $g_{\star106}\equiv g_{\star,{\rm rh}}/106.75$ and 
	\begin{equation}
		f_{c}\equiv\frac{18.9\ \text{GHz}}{g_{\star106}^{1/3}}\thinspace.\label{eq:O1-cata}
	\end{equation}
	The spectrum peaks at  $f\sim 3 f_c$, corresponding to $k\sim 3 T_{\rm rh}$ in Fig.~\ref{fig:cata}.
	To estimate the peak of the spectrum, one can take $\Upsilon_{7/5}(3)\sim 0.1$. The approximations in this section are helpful for understanding the behavior of the radiation-catalyzed GWs. We would like to stress that the GW spectrum presented in the next section is based on full numerical computations.
	
	Comparing Eq.~\eqref{eq:ogw-cata} to Eq.~\eqref{eq:ogw-bremJ}, 
	we see that the radiation-catalyzed channel can generate a much higher GW spectrum than the bremsstrahlung channel under certain circumstances. We  will also show this more explicitly in the next section.
	
	\subsubsection{GWs from radiation-radiation scattering \label{subsec:RR}}
	
	As has been illustrated in Fig.~\ref{fig:schematic}, during reheating
	and after reheating, gravitons can be produced from radiation-radiation
	scattering. The exact production rate depends on how thermal species
	in the thermal bath interact with each other. For the SM thermal bath
	in the radiation-dominated universe (corresponding to the epoch indicated
	by $R\in\text{SM}$ in Fig.~\ref{fig:schematic}), this production
	has been calculated in Refs.~\cite{Ghiglieri:2015nfa,Ghiglieri:2020mhm}.
	For more extensive studies on the production of gravitons from the
	SM or BSM thermal bath, see~\cite{Ringwald:2020ist,Klose:2022knn,Ringwald:2022xif,Klose:2022rxh,Ghiglieri:2022rfp,Muia:2023wru,Drewes:2023oxg,Bernal:2024jim}. 
	
	Since the early universe at a very high temperature (well above the
	electroweak scale) could be dominated by very different particle contents
	(e.g., $SO(10)$ plasma), we remain agnostic regarding the particle
	physics models and consider a simplified model consisting of $N_{f}$
	fermions ($\psi_{1}$, $\psi_{2}$, $\psi_{3}$, ..., $\psi_{N_{f}}$)
	and $N_{g}$ gauge bosons ($A_{1}$, $A_{2}$, $A_{3}$, ..., $A_{N_{g}}$).
	Each gauge boson is coupled to every fermion with a universal gauge
	coupling $g$, i.e., 
	\begin{equation}
		{\cal L}_{R}\supset\sum_{i,j}g\overline{\psi_{i}}A_{j}^{\mu}\gamma_{\mu}\psi_{i}\thinspace.\label{eq:gauge-int}
	\end{equation}
	Self-interactions of gauge bosons, which would be present in non-Abelian
	gauge theories, are neglected for simplicity. Given that most interactions
	in the SM and many BSM models are gauge interactions, the simplified
	model is expected to capture the essential characteristics of graviton
	production in the SM and BSM thermal bath. 
	
	In the simplified model, the leading-order production of gravitons
	arises from the processes $\psi\overline{\psi}\to hA$ and $A\psi\to h\psi$---see
	the last two diagrams in Fig.~\ref{fig:Rep-Feyn}. The spin/polarization
	averaged matrix elements for these two processes read
	\begin{align}
		|\overline{\mathcal{M}}_{\psi\overline{\psi}\to hA}|^{2} & =\frac{g^{2}}{4M_{P}^{2}}\frac{t^{2}+u^{2}}{s}\thinspace,\label{eq:M2-scat-1}\\
		|\overline{\mathcal{M}}_{A\psi\to h\psi}|^{2} & =\frac{g^{2}}{4M_{P}^{2}}\frac{u^{2}+s^{2}}{t}\thinspace,\label{eq:M2-scat-2}
	\end{align}
	where $s$, $t$, and $u$ are the Mandelstam variables of these two-to-two
	scattering processes. 
	
	Note that $|\overline{\mathcal{M}}_{A\psi\to h\psi}|^{2}$ would be
	divergent in the limit of $t\to0$. This divergence is regulated by
	Debye-H\"uckel screening---see Appendix~\ref{sec:Calc-collision}
	for details. After the screening effect is taken into account, we
	obtain the following collision term: 
	\begin{align}
		\mathcal{C}_{h}\simeq\frac{N_{f}N_{g}g^{2}}{\pi^{3}M_{P}^{2}}T^{3}e^{-\omega/T}\left[\frac{1}{12}+\frac{1}{8}G\left(\frac{\omega}{\kappa}\right)\right],\label{eq:Ch_psipsi}
	\end{align}
	where $\kappa=g\sqrt{n_{\psi}/T}$ is the Debye-H\"uckel screening
	scale and 
	\begin{equation}
		G(x)\equiv-\frac{3}{2}-\frac{1}{4x^{2}}+\left(2+\frac{1}{2x^{2}}+\frac{1}{16x^{4}}\right)\ln\left(1+4x^{2}\right).\label{eq:G}
	\end{equation}
	In Eq.~\eqref{eq:Ch_psipsi}, the first and second terms correspond
	to the contributions of $\psi\overline{\psi}\to hA$ and $A\psi\to h\psi$,
	respectively. Note that $G(x)$ in Eq.~\eqref{eq:G} is always positive,
	despite its first two negative terms. In particular, when $x$ is
	small, the logarithmic term approaches $3/2+1/(4x^{2})+16x^{2}/3+{\cal O}(x^{4})$,
	implying that $\lim_{x\to0}G(x)=16x^{2}/3$. In general, the $G$-term
	in Eq.~\eqref{eq:Ch_psipsi} is greater than the other term, as long
	as $\omega/\kappa\gtrsim0.4$. 
	
	\begin{figure}
		\centering 
		
		\includegraphics[width=0.6\textwidth]{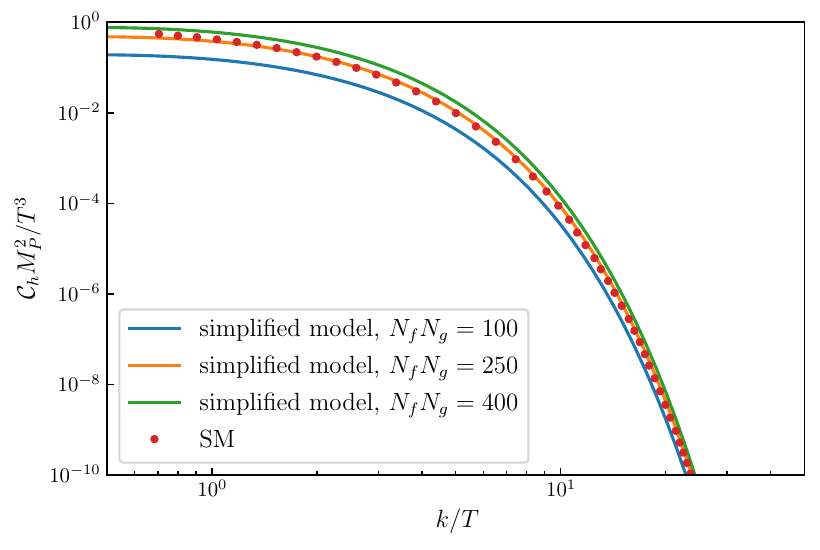} \caption{Comparison of our thermal graviton production rate in the simplified
			model (solid lines) with the SM calculation (red dots). The solid
			lines are produced using Eq.~(\ref{eq:Ch_psipsi}), with $g^{2}=0.1$
			and $N_{f}N_{g}$ specified in the plot. The SM values are taken from
			Ref.~\cite{Ringwald:2020ist}. \label{fig:compare-to-SM} }
		
	\end{figure}
	
	In Fig.~\ref{fig:compare-to-SM}, we compare the production rate
	in Eq.\ (\ref{eq:Ch_psipsi}) with the SM production rate, which
	has been previously calculated in Refs.~\cite{Ghiglieri:2015nfa,Ghiglieri:2020mhm,Ringwald:2020ist}.
	The SM values shown in this figure are taken from Figure 1 of Ref.~\cite{Ringwald:2020ist},
	where a dimensionless quantity $\hat{\eta}$ was used to present the
	result. According to Eq.~(2.4) in Ref.~\cite{Ringwald:2020ist},
	we use $\hat{\eta}=\frac{M_{P}^{2}}{4T^{4}}k\mathcal{C}_{h}$ to recast
	the result. As is shown in Fig.~\ref{fig:compare-to-SM}, the SM
	results can be well  approximated by our simplified model with $N_{f}N_{g}=250$
	and $g^{2}=0.1$. 
	
	From the collision term to $f_{h}$, the calculation is similar to
	that in Sec.~\ref{subsec:catalyzed} and yields
	\begin{equation}
		f_{h}(a_{{\rm rh}},k)\approx\frac{N_{f}N_{g}g^{2}T_{{\rm rh}}^{3}}{15\pi^{3}H_{I}M_{P}^{2}}\frac{2+3\overline{G}}{r_{a}^{3/2}}\left(\frac{k}{T_{\text{rh}}}\right)^{\frac{3}{5}}\Upsilon_{-\frac{3}{5}}\left(\frac{k}{T_{\text{rh}}}\right),\label{eq:f-RR}
	\end{equation} 
	where $\overline{G}$ denotes the thermal average of the $G$ factor
	in Eq.~\eqref{eq:G}. Its specific value depends on the Debye-H\"uckel
	screening scale $\kappa$ and hence on $g^{2}$. However, due to the
	logarithmic function in $G$, its dependence on $g^{2}$ is weak.
	For $g^{2}$ varying from $\sim10^{-2}$ to $\sim1$, $\overline{G}$
	only varies from $\sim10$ to $\sim20$. So in practice, we recommend
	taking $\overline{G}\approx15$ for SM-like thermal plasma. 
	
	Using Eq.~\eqref{eq:f-RR}, we obtain the following the GW spectrum
	\begin{equation}
		\Omega_{{\rm GW}}h^{2}(f)\simeq6.5\times10^{-17}\left(\frac{N_{f}N_{g}g^{2}}{25}\right)\left(\frac{\Gamma_{\phi}}{10^{-5}M_{P}}\right)^{\frac{1}{2}}\left(\frac{f}{\text{GHz}}\right)^{\frac{23}{5}}\frac{\Upsilon_{-\frac{3}{5}}(f/f_c)}{g_{\star106}^{11/20}}\thinspace,\label{eq:ogw-RR}
	\end{equation}
	where the last part containing the $\Upsilon$ function and $g_{\star106}$ is typically around ${\cal O}(1)$ at the peak of the spectrum.  
	Note that the $\Upsilon_{-\frac{3}{5}}$ function asymptotically behaves as $\frac{5}{3}  (f/f_c)^{-3/5}$ for $f\ll f_c$ and $r_a\to 0$. Therefore, in the low-$f$ regime, Eq.~\eqref{eq:ogw-RR} implies a power law of $f^{23/5}\cdot f^{-3/5}=f^4$, provided that $r_a$ is sufficiently small.
	
	\section{Results}\label{sec:results}
	In this section, we present the complete gravitational wave (GW) spectra generated in our framework spanning over the period from inflation through reheating and eventually to the radiation-dominated era. For the post-inflationary GW spectra, we numerically solve the Boltzmann equation for the graviton phase space distribution function in each scenario, along with the background evolution equations, Eq.~\eqref{eq:rhophi} and Eq.~\eqref{eq:rhoR}. The computation begins at the onset of reheating and extends through several e-folds beyond its completion. This approach ensures that our result fully includes the contributions from the epochs before and after reheating.
	\begin{figure}
		\centering
		\includegraphics[width=0.49\textwidth]{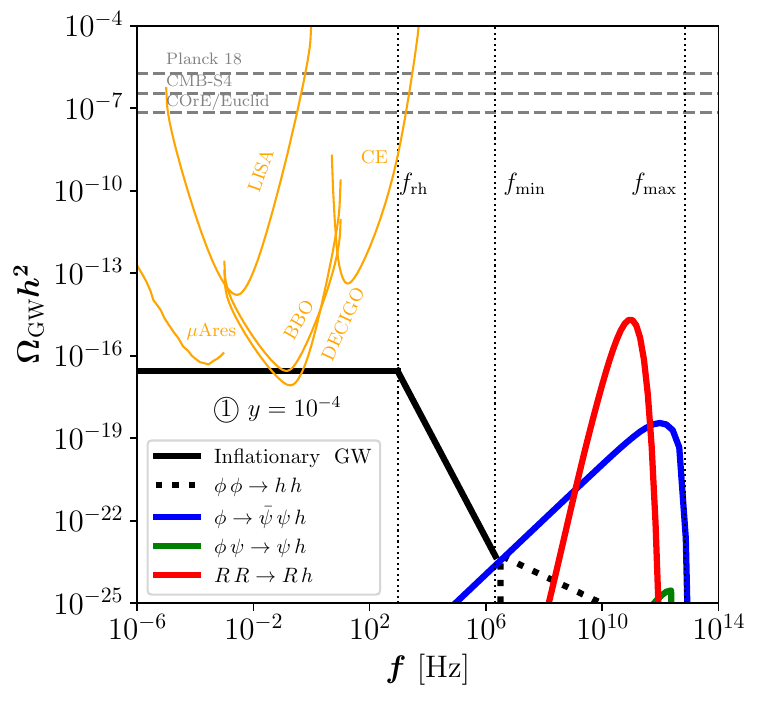}
		\includegraphics[width=0.49\textwidth]{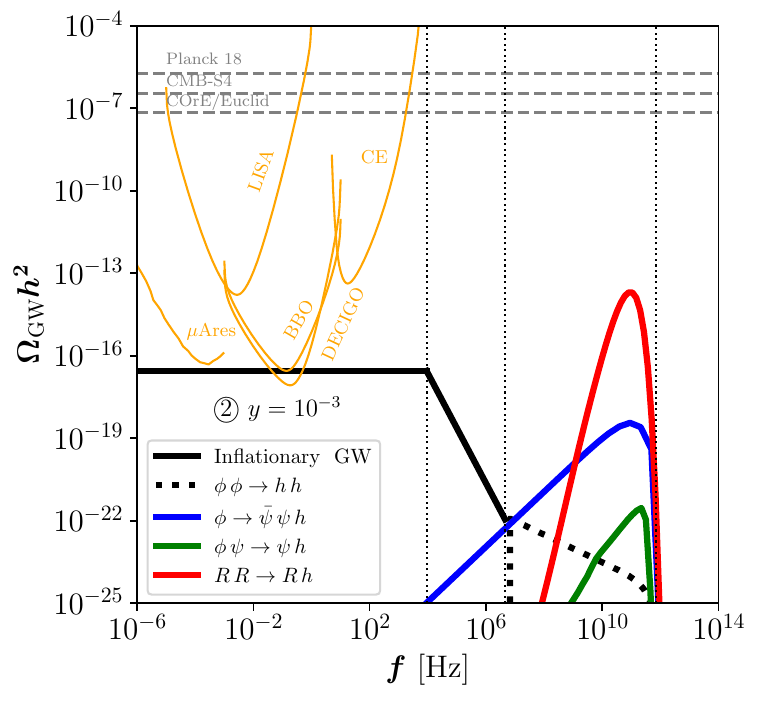}
		\includegraphics[width=0.49\textwidth]{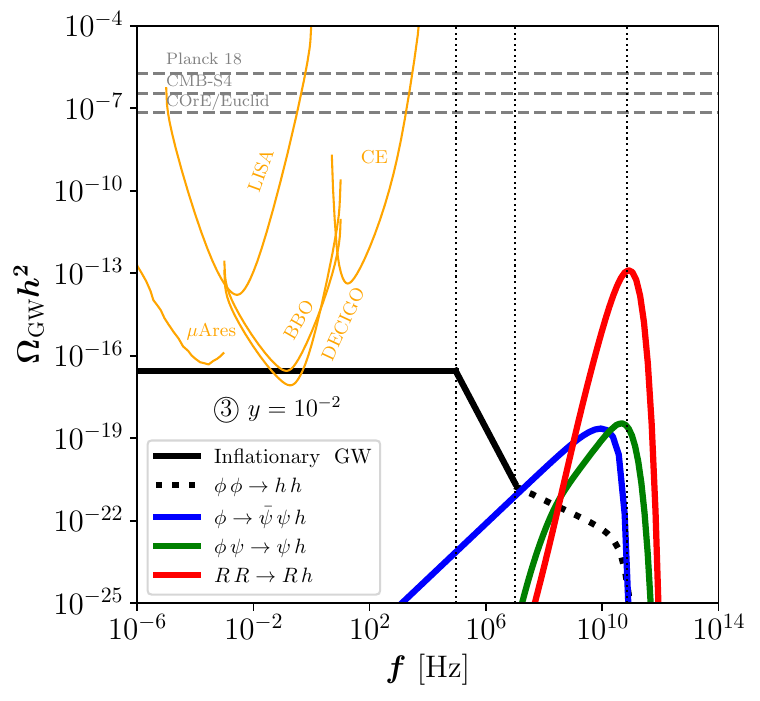}
		\includegraphics[width=0.49\textwidth]{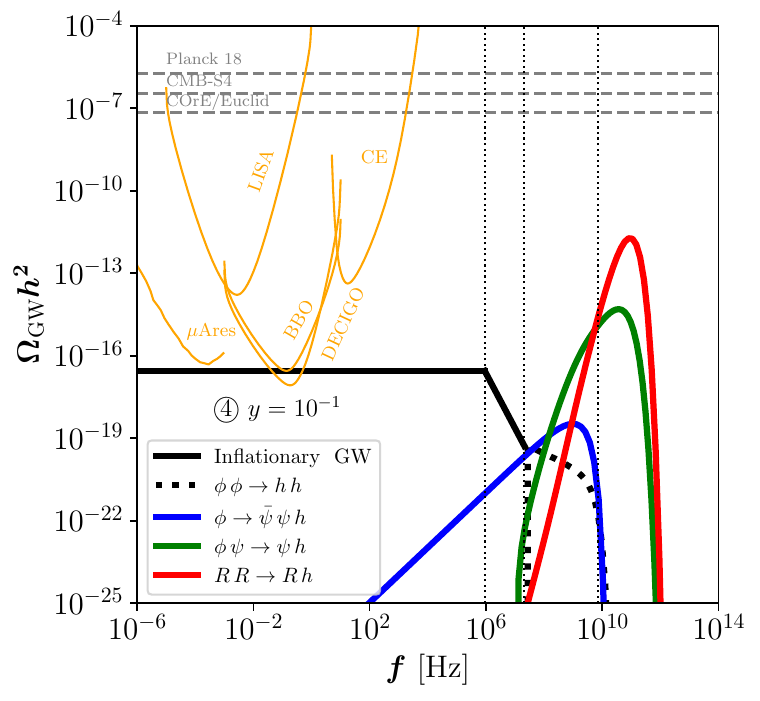}
		\caption{Comparison of the different sources of GWs in three benchmarks with 
			\tcircled{1} $y = 10^{-4}$, \tcircled{2} $y = 10^{-3}$,    \tcircled{3} $y = 10^{-2}$,  and \tcircled{4} $y = 10^{-1}$. 
			Other parameters are commonly set at    
			$g^2 =10^{-1}$, $m_\phi = 10^{13}$~GeV, $r=0.02$, $H_I \simeq 10^{13}$~GeV, and $ N_f N_g =250$. 
			The three vertical lines correspond to $f=f_\text{rh}$ and $f=f_I$, and $f=f_\text{max}$, given by Eqs.~\eqref{eq:frh}, \eqref{eq:fI}, and \eqref{eq:fmax}, respectively.  }
		\label{fig:GW}
	\end{figure} 

	The results are shown in Fig.~\ref{fig:GW}, along with the projected sensitivity curves of future detectors, including the Big Bang Observer (BBO)~\cite{Crowder:2005nr, Corbin:2005ny}, DECIGO~\cite{Seto:2001qf, Kudoh:2005as}, LISA~\cite{LISA:2017pwj}, $\mu$Ares~\cite{Sesana:2019vho}, the Cosmic Explorer (CE)~\cite{Reitze:2019iox}, and the Einstein Telescope (ET)~\cite{Hild:2010id,Punturo:2010zz, Sathyaprakash:2012jk,ET:2019dnz}. These sensitivity curves are shown in yellow color. The energy stored in GWs behaves like dark radiation and contributes to the effective number of neutrino species, $N_\text{eff}$.\footnote{To constrain this contribution, we use the bound $\ogw h^2 \lesssim 5.6 \times 10^{-6} \Delta N_{\text{eff}}$, which applies to the total energy density integrated over logarithmic frequency~\cite{Caprini:2018mtu}.} 
	The Planck 2018 results provide a 95\% confidence level measurement of $N_{\text{eff}} = 2.99 \pm 0.34$~\cite{Planck:2018vyg}.  The proposed CMB-S4 experiment is projected to reach a sensitivity of $\Delta N_{\text{eff}}\lesssim 0.06$~\cite{Abazajian:2019eic}. 
	Future surveys, such as COrE~\cite{COrE:2011bfs} and Euclid~\cite{EUCLID:2011zbd}, are expected to tighten this constraint to $\Delta N_{\text{eff}}\lesssim 0.013$ at the $2\sigma$ level. These limits are illustrated by horizontal dashed lines in Fig.~\ref{fig:GW}.

	We consider four benchmark scenarios characterized by different values of the coupling $y$ and corresponding reheating temperatures $T_{\text{rh}}$:  $\circled{1}$  $y = 10^{-4}$ with $T_{\text{rh}} \simeq 10^{10}\,\text{GeV}$ (top left), $\circled{2}$ $y = 10^{-3}$ with $T_{\text{rh}} \simeq 10^{11}\,\text{GeV}$ (top right),  $\circled{3}$  $y = 10^{-2}$ with $T_{\text{rh}} \simeq 10^{12}\,\text{GeV}$ (bottom left),  and  $\circled{4}$  $y = 10^{-1}$ with $T_{\text{rh}} \simeq 10^{13}\,\text{GeV}$ (bottom right).  In all cases, we fix the remaining model parameters to $g^2 = 10^{-1}$, $N_g N_f =250$, $m_\phi = 10^{13}\,\text{GeV}$, $r = 0.02$, and $H_I \simeq 10^{13}\,\text{GeV}$ (cf. Eq.~\eqref{eq:Hinf}). We emphasize that once the coupling $y$, the inflaton mass, and the inflationary Hubble scale $H_I$ are specified, the reheating temperature is determined accordingly by solve the background Eq.~\eqref{eq:rhophi} and Eq.~\eqref{eq:rhoR}. 
	
	Below we list and discuss some features of the GW spectra under consideration.
	\begin{itemize}
		\item The solid black line represents the inflationary gravitational wave spectrum generated during inflation. The flat portion corresponds to tensor modes that re-enter the horizon during the radiation-dominated era, while the subsequent $f^{-2}$ decline reflects modes that re-enter during the reheating phase, as is formulated in the second line of Eq.~\eqref{eq:ogw-1}.
		\item The black dotted line represents the high-frequency tail of tensor modes, which never exit the horizon but can still be excited during reheating via inflaton annihilation. According to Eq.~\eqref{eq:ogw_inflaton_inflaton}, this component of the spectrum decreases as $f^{-1/2}$  until it reaches the cutoff frequency $f_{\text{max}}$ given by Eq.~\eqref{eq:fmax}, beyond which the signal is exponentially suppressed due to inflaton decay.
		\item The solid blue line represents the GW spectrum generated from bremsstrahlung. The signal peaks when the energy of the emitted graviton reaches its kinematic threshold, corresponding to $\omega = m_\phi/2$ at the end of reheating. This sets the peak frequency at $f_{\text{peak}} \simeq f_{\text{max}}/2$. In the regime $f < f_{\text{peak}}$, the GW spectrum scales as $\Omega_{\text{GW}} \propto f$ (cf. Eq.~\eqref{eq:ogw-bremJ}), while for $f \gtrsim f_{\text{peak}}$, the spectrum is exponentially suppressed due to the rapid depletion of the inflaton number density.
		\item  The solid green curve represents  the GW spectrum generated from the radiation-catalyzed scattering process $\phi \psi \to \psi h$. The produced gravitons carry energy $\omega > m_\phi /2$, which implies a lower limit on the frequency, given by $f \simeq f_{\text{min}}/2$, where $f_{\text{min}}$ is defined in Eq.~\eqref{eq:fmin}. At low temperatures, the Boltzmann suppression can be neglected, and the collision term in Eq.~\eqref{eq:Ch-catalyze} peaks at $\omega \simeq m_\phi$, corresponding to a GW frequency $f \simeq f_{\text{max}}$. 
		At higher temperatures, gravitons are produced by more energetic radiation, leading to a shift of the peak frequency toward that of the thermal GW spectrum, which will be discussed in the next bullet point.

		%
		\item The solid red line represents the GW spectrum arising from pure thermal scatterings via the processes $\bar{\psi} \psi \to A\, h$ and $A \psi \to \psi h$, with $N_f N_g = 250$. As discussed in the previous section, this scenario can reproduce the GW signal from Standard Model (SM) thermal scatterings. Similar to the SM case, the spectrum exhibits a peak at $f_{\text{peak}} \simeq 100~\text{GHz}$ \cite{Ringwald:2020ist}. 
		In the low-frequency regime, the spectrum is dominated by the $\bar{\psi} \psi \to A\, h$ process, leading to a scaling of $\Omega_{\text{GW}} \propto f^4$---see the discussion below  Eq.~\eqref{eq:ogw-RR}. 
		At higher frequencies close to the peak, the spectrum receives the dominant contribution from  $A \psi \to \psi h$. 
		Most gravitons are produced with energies on the order of the temperature at the time of production, while excessively high energy gravitons are exponentially suppressed due to the Boltzmann factor $e^{-\omega/T}$ in Eq.~\eqref{eq:Ch_psipsi}.
	\end{itemize}
	By comparing the spectra in Fig.~\ref{fig:GW} for the four benchmark model parameters, we find that the inflationary GW signal dominates in the regime $f < f_I$, where $f= f_I$ is depicted as the second vertical gray line. For $f > f_I$, the contributions from the other four sources begin to interplay. In general, the bremsstrahlung channel $\phi \to \bar{\psi} \psi h$ dominates over the pure inflaton annihilation process $\phi \phi \to hh$ 
	when $m_\phi \gtrsim T_{\text{rh}}$, except in the high-frequency tail and a narrow low-frequency region. Compared to the $1 \to 3$ bremsstrahlung process, the radiation-catalyzed $2 \to 2$ scatterings involving the inflaton and its daughter fields become significant when $T_{\text{rh}}$ is not much smaller than $m_\phi$, as shown in the lower panels of Fig.~\ref{fig:GW}.

	Our main results concern the GW spectrum sourced by thermal scatterings. Since pure thermal scatterings correspond to a UV freeze-in process, the production rate increases with temperature. This explains why the peak amplitude of the pure thermal GW spectrum grows with larger values of $y$ or $\Trh$; see the peaks of the red curves in Fig.~\ref{fig:GW}.  We note that the peak can reach $\Omega_{\text{GW}} h^2 \sim \mathcal{O}(10^{-12})$ for a reheating temperature of $\Trh \simeq 10^{13}\,\text{GeV}$, as shown in the bottom right panel of Fig.~\ref{fig:GW}. More interestingly, we find that the pure thermal scattering channel can dominate GW production even in scenarios with $\Trh \ll m_\phi$, and even with a single species of fermions and gauge bosons, i.e., $N_f N_g = 1$. This can be seen in the upper left panel of Fig.~\ref{fig:GW}, where the thermal GW peak reaches approximately $\mathcal{O}(10^{-15}) / 250 \sim \mathcal{O}(10^{-17})$ for $N_f N_g = 1$.  When additional species are included—i.e., for large $N_f N_g$—the amplitude of the thermal GW signal increases further, as shown in Eq.~\eqref{eq:ogw-RR}.
	
	\section{Conclusions} \label{sec:conclusion}
	In this work, we aim to provide a full-spectrum analysis of GW production
	within a generic framework that assumes slow-roll inflation followed
	by a reheating phase, during which cold inflatons gradually decay
	into radiation, ultimately driving the universe into a radiation-dominated
	era. By employing the Boltzmann equation of the graviton phase space
	distribution function, we systematically compute the GW spectra generated
	by $(i)$ pure inflaton annihilation, $(ii)$ graviton bremsstrahlung
	from inflaton decay, $(iii)$ radiation-catalyzed inflaton-graviton
	conversion, and $(iv)$ scattering among fully thermalized radiation
	particles. For each channel, we obtain the corresponding collision
	term with the calculation presented in great detail in the appendixes,
	and solve the Boltzmann equation numerically to get the GW spectrum.
	Moreover, we derive accurate and simple analytical formulae for these
	GW spectra---see Eqs.~\eqref{eq:ogw_inflaton_inflaton}, \eqref{eq:ogw-bremJ},
	\eqref{eq:ogw-cata}, and \eqref{eq:ogw-RR}---which we believe would
	be useful to relevant studies. 
	
	Our main results are illustrated in Fig.~\ref{fig:GW}, where we compare the inflationary GW spectrum with contributions from reheating-era sources. We find that inflationary GWs dominate the low-frequency regime, $f < f_I$, where $f_I$ denotes the frequency of modes re-entering the horizon at the onset of reheating. At higher frequencies, the GW spectrum is shaped by the interplay of reheating-era sources. Among them, the dominant contribution arises from $(iv)$—even when the reheating temperature is much lower than the inflaton mass.
	
	It is worth mentioning that for GWs produced in $(iv)$,  we have considered a simplified model consisting solely of $N_f$ fermions and $N_g$ gauge bosons. Remarkably, we find that our result obtained in this simplified model with $N_f N_g \sim 250$ can approximate the SM thermal production rate of gravitons very well---see Fig.~\ref{fig:compare-to-SM}. We believe that our result in the simplified model can be used to facilitate the calculation of GW production in various BSM thermal plasma. 
	
	In summary, we have presented the first full-spectrum analysis of inflationary and post-inflationary GWs. Our work demonstrates that post-inflationary physics can generate high-frequency GWs with rich structures, which, if observable by any means, could reveal crucial information about inflation and post-inflationary reheating.

	\acknowledgments
	We thank Nicolás Bernal for his helpful comments on the draft and Marco Drewes for valuable discussions.  YX would like to thank Shi Pi for discussions  during his visit at the Institute of Theoretical Physics, Chinese Academy of Sciences in 2024. The work of XJX is supported in part by the National Natural Science Foundation of China (NSFC) under grant No.~12141501 and also by the CAS Project for Young Scientists in Basic Research (YSBR-099).
	YX has received support from the Cluster of Excellence ``Precision Physics, Fundamental Interactions, and Structure of Matter'' (PRISMA$^+$ EXC 2118/1) funded by the Deutsche Forschungsgemeinschaft (DFG, German Research Foundation) within the German Excellence Strategy (Project No. 390831469). 
	
	\appendix
	
	\section{Calculation of horizon crossing\label{sec:hc}}
	
	Solving the horizon crossing scale factor $a_{{\rm hc}}$ in Eq.~\eqref{eq:ah-k}
	requires the explicit form of $H(a)$  during the radiation-dominated
	and reheating epochs. 
	
	Let us first consider the radiation-dominated epoch, in which the $T$-$a$ relation is given by Eq.~\eqref{eq:T-a}. 
	
	Eq.~\eqref{eq:T-a} implies
	that the temperature at matter-radiation equality is
	\begin{equation}
		\Teq\simeq T_{0}\frac{a_{0}}{a_{{\rm eq}}}\approx0.8\ {\rm eV}\thinspace.\label{eq:Teq}
	\end{equation}
	Using Eq.~\eqref{eq:T-a}, one can express $\rho_{R}$ and $H=\sqrt{\rho_{R}/(3M_{P}^{2})}$
	as functions of $a$:
	\begin{align}
		\rho_{R}(a) & =\left(\frac{a_{0}}{a}\right)^{4}\frac{\pi^{2}g_{\star}}{30}T_{0}^{4}\left(\frac{g_{\star s,0}}{g_{\star s}}\right)^{4/3},\ \ {\rm for}\ \ a\geq a_{{\rm rh}}\thinspace,\label{eq:rho-a-RD}\\
		H(a) & =\frac{\pi T_{0}^{2}}{3M_{P}}\left(\frac{a_{0}}{a}\right)^{2}\sqrt{\frac{g_{\star}}{10}}\left(\frac{g_{\star s,0}}{g_{\star s}}\right)^{2/3},\ \ {\rm for}\ \ a\in[a_{{\rm rh}},\ a_{{\rm eq}}]\thinspace.\label{eq:H-a-RD}
	\end{align}
	Substituting Eq.~\eqref{eq:H-a-RD} into Eq.~\eqref{eq:ah-k}, we
	obtain the first case in Eq.~\eqref{eq:ahc}.
	
	Next, we turn to the reheating epoch, in which the expansion is dominantly
	driven by matter, $\rho_{\phi}\propto a^{-3}$. Its specific form
	can be inferred from the end of this epoch, at which the energy density
	is approximately $\rho_{R}(a_{{\rm rh}})$: 
	\begin{equation}
		\rho_{\phi}\approx\left(\frac{a_{{\rm rh}}}{a}\right)^{3}\rho_{R}(a_{{\rm rh}})\approx\left(\frac{a_{{\rm rh}}}{a}\right)^{3}\frac{\pi^{2}g_{\star,\text{rh}}}{30}T_{{\rm rh}}^{4}\thinspace,\ \ {\rm for}\ \ a\in[a_{I},\ a_{{\rm rh}}]\thinspace.\label{eq:rho-a-RH}
	\end{equation}
	Therefore, the Hubble parameter is given by
	\begin{equation}
		H(a)=\frac{\pi T_{{\rm rh}}^{2}}{3M_{P}}\left(\frac{a_{{\rm rh}}}{a}\right)^{\frac{3}{2}}\sqrt{\frac{g_{\star,{\rm rh}}}{10}}\thinspace,\ \ {\rm for}\ \ a\in[a_{I},\ a_{{\rm rh}}]\thinspace.\label{eq:H-a-RH}
	\end{equation}
	Substituting Eq.~\eqref{eq:H-a-RH} into Eq.~\eqref{eq:ah-k}, we
	obtain the second case in Eq.~\eqref{eq:ahc}.
	
	The frequency $f_{I}$ in Eq.~\eqref{eq:fI} is obtained by expressing
	$a_{I}$ in terms of $T_{{\rm rh}}$:
	\begin{equation}
		a_{I}\approx a_{{\rm rh}}\left(\frac{2\Gamma_{\phi}}{3H_{I}}\right)^{\frac{2}{3}}\approx a_{{\rm rh}}\left(\frac{\pi^{2}g_{\star,\text{rh}}T_{{\rm rh}}^{4}}{90H_{I}^{2}M_{P}^{2}}\right)^{\frac{1}{3}}\approx\left(\frac{\pi^{2}g_{\star,\text{rh}}T_{{\rm rh}}T_{0}^{3}}{90H_{I}^{2}M_{P}^{2}}\cdot\frac{g_{\star s,0}}{g_{\star s,{\rm rh}}}\right)^{\frac{1}{3}},\label{eq:aI-Trh}
	\end{equation}
	where the first, second, and third steps have used Eqs.~\eqref{eq:a_rh},
	\eqref{eq:T_rh}, and \eqref{eq:T-a},  respectively.

	\section{Calculation of matrix elements\label{sec:M2}}
	
	In this appendix,  we present the detailed calculation of the matrix
	elements used in this work. We shall mention here that some of the
	matrix elements have already been calculated in the literature~\cite{Choi:2024ilx,Barman:2023ymn,Xu:2024fjl,Bernal:2025lxp}.
	Only the matrix elements for $\psi\overline{\psi}\to hA$ and $A\psi\to h\psi$
	are new. Nevertheless, we believe it is useful to include all the
	matrix elements in a self-contained manner, with unified conventions
	and notations. 
	
	\subsection{Gravitational Polarization Summation (GPS)}
	
	For spin-2 gravitons with four-momentum $q^{\mu}=\left(\omega,\ \vec{q}\right)$,
	the polarization tensor $\epsilon^{\mu\nu}$ satisfies the following
	conditions: 
	\begin{equation}
		\begin{aligned}\epsilon^{i\mu\nu}=\epsilon^{i\nu\mu} & \text{ symmetric }\\
			q_{\mu}\epsilon^{i\mu\nu}=0 & \text{ transverse }\\
			\eta_{\mu\nu}\epsilon^{i\mu\nu}=0 & \text{ traceless }\\
			\epsilon^{i\mu\nu}\epsilon_{\mu\nu}^{*j}=\delta^{ij} & \text{ orthonormal }
		\end{aligned}
		\thinspace,\label{eq:A}
	\end{equation}
	where $i,j$ are polarization indices. Introducing the auxiliary vector
	$\bar{q}^{\mu}=\left(\omega,\ -\vec{q}\right)$, the gravitational
	polarization summation (GPS) reads~\cite{Barman:2023ymn}:
	\begin{equation}
		\mathrm{GPS}_{\mu\nu,\alpha\beta}\equiv\sum_{\text{pol }}\epsilon_{\mu\nu}^{*}\epsilon_{\alpha\beta}=\frac{1}{2}\left(\hat{\eta}_{\mu\alpha}\hat{\eta}_{\nu\beta}+\hat{\eta}_{\mu\beta}\hat{\eta}_{\nu\alpha}-\hat{\eta}_{\mu\nu}\hat{\eta}_{\alpha\beta}\right),\label{eq:A-1}
	\end{equation}
	where
	\begin{equation}
		\hat{\eta}_{\mu\nu}\equiv\eta_{\mu\nu}-\frac{q_{\mu}\bar{q}_{\nu}+\bar{q}_{\mu}q_{\nu}}{q\cdot\bar{q}}=\left(\begin{array}{cccc}
			0 & 0 & 0 & 0\\
			0 & \frac{q_{1}^{2}}{\omega^{2}}-1 & \frac{q_{1}q_{2}}{\omega^{2}} & \frac{q_{1}q_{3}}{\omega^{2}}\\
			0 & \frac{q_{1}q_{2}}{\omega^{2}} & \frac{q_{2}^{2}}{\omega^{2}}-1 & \frac{q_{2}q_{3}}{\omega^{2}}\\
			0 & \frac{q_{1}q_{3}}{\omega^{2}} & \frac{q_{2}q_{3}}{\omega^{2}} & \frac{q_{3}^{2}}{\omega^{2}}-1
		\end{array}\right)\thinspace.\label{eq:A-2}
	\end{equation}
	Due to the vanishing
	temporal components of $\hat{\eta}_{\mu\nu}$,  the GPS satisfies
	\begin{equation}
		\mathrm{GPS}_{\mu\nu,\alpha\beta}X^{\mu}=0\ \ \text{for}\ \forall\ X^{\mu}=\left(X^{0},\ 0,\ 0,\ 0\right).\label{eq:A-3}
	\end{equation}
	This feature allows us to significantly simplify calculations involving
	GPS and non-relativistic particles.

	\subsection{$\phi\phi\rightarrow hh$}
	
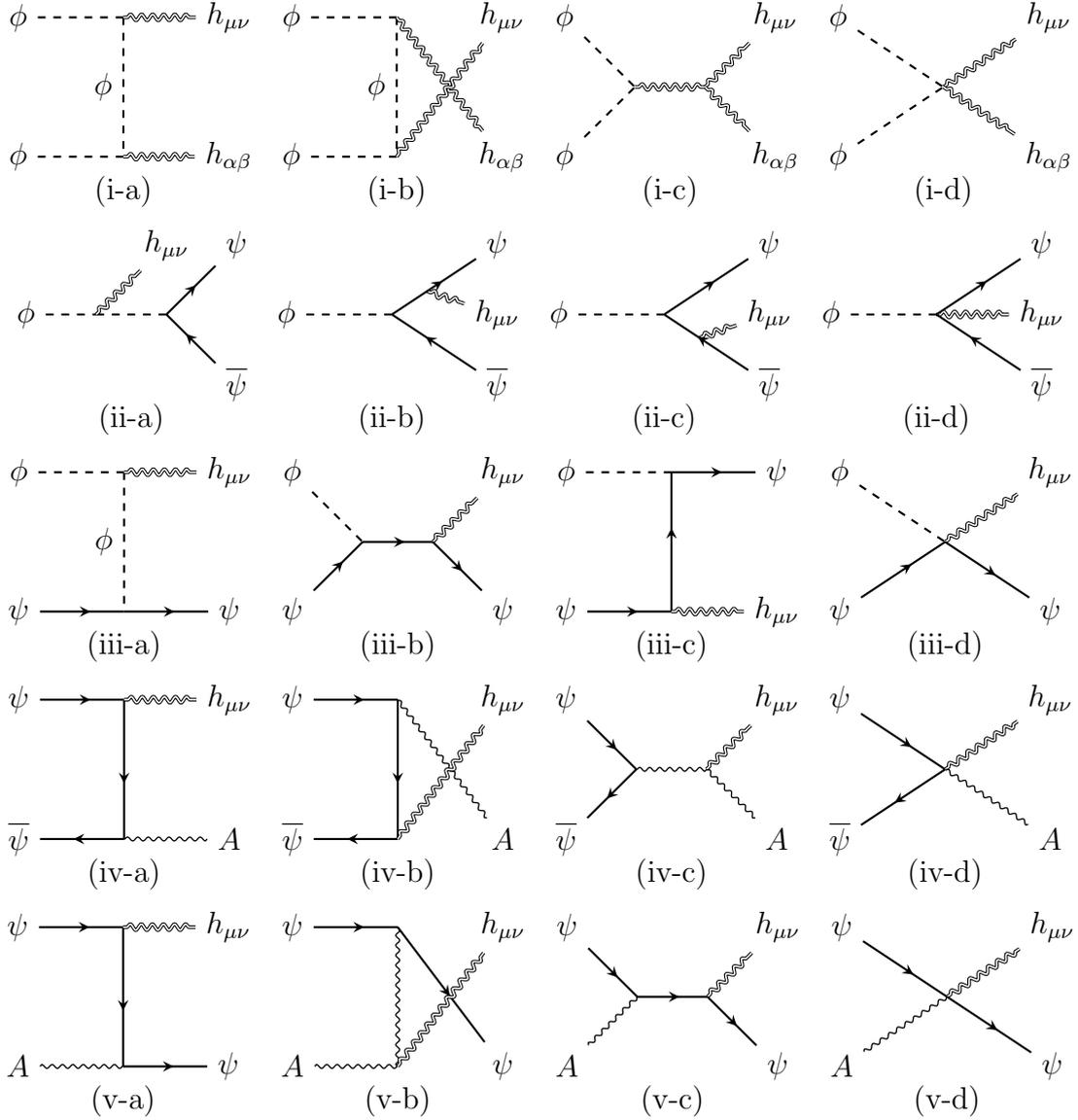
\begin{figure}
		\centering\begin{tikzpicture}[scale=0.95]
			\begin{feynhand}
				\vertex (1) at (-1.5, 1) {$\phi$};
				\vertex (2) at (-1.5, -1) {$\phi$};
				\vertex (u) at (0, 1);
				\vertex (d) at (0, -1);
				\vertex (3) at (1.5, 1) {$h_{\mu\nu}$};
				\vertex (4) at (1.5, -1) {$h_{\alpha\beta}$};
				\vertex () at (0, -1.5) {(i-a)};
				
				\propag [inflaton] (1) to (u);
				\propag [inflaton] (2) to (d);
				\propag [inflaton] (u) to [edge label'=$\phi$] (d);
				\propag [graviton] (u) to (3);
				\propag [graviton] (d) to (4);
			\end{feynhand}
		\end{tikzpicture}
		\begin{tikzpicture}[scale=0.95]
			\begin{feynhand}
				\vertex (1) at (-1.5, 1) {$\phi$};
				\vertex (2) at (-1.5, -1) {$\phi$};
				\vertex (u) at (0, 1);
				\vertex (d) at (0, -1);
				\vertex (3) at (1.5, 1) {$h_{\mu\nu}$};
				\vertex (4) at (1.5, -1) {$h_{\alpha\beta}$};
				\vertex () at (0, -1.5) {(i-b)};
				
				\propag [inflaton] (1) to (u);
				\propag [inflaton] (2) to (d);
				\propag [inflaton] (u) to [edge label'=$\phi$] (d);
				\propag [graviton] (u) to (4);
				\propag [graviton] (d) to (3);
			\end{feynhand}
		\end{tikzpicture}
		\begin{tikzpicture}[scale=0.95]
			\begin{feynhand}
				\vertex (1) at (-1.5, 1) {$\phi$};
				\vertex (2) at (-1.5, -1) {$\phi$};
				\vertex (L) at (-0.5, 0);
				\vertex (R) at (0.5, 0);
				\vertex (3) at (1.5, 1) {$h_{\mu\nu}$};
				\vertex (4) at (1.5, -1) {$h_{\alpha\beta}$};
				\vertex () at (0, -1.5) {(i-c)};
				
				\propag [inflaton] (1) to (L);
				\propag [inflaton] (2) to (L);
				\propag [graviton]  (L) to  (R);
				\propag [graviton] (R) to (3);
				\propag [graviton] (R) to (4);
			\end{feynhand}
		\end{tikzpicture}
		\begin{tikzpicture}[scale=0.95]
			\begin{feynhand}
				\vertex (1) at (-1.5, 1) {$\phi$};
				\vertex (2) at (-1.5, -1) {$\phi$};
				\vertex (M) at (0, 0);
				\vertex (3) at (1.5, 1) {$h_{\mu\nu}$};
				\vertex (4) at (1.5, -1) {$h_{\alpha\beta}$};
				\vertex () at (0, -1.5) {(i-d)};
				
				\propag [inflaton] (1) to (M);
				\propag [inflaton] (2) to (M);
				\propag [graviton] (M) to (3);
				\propag [graviton] (M) to (4);
			\end{feynhand}
		\end{tikzpicture}
		
		
		\begin{tikzpicture}[scale=0.95]
			\begin{feynhand}
				\vertex (i) at (-1.5, 0) {$\phi$};
				\vertex (L) at (-0.5, 0);
				\vertex (R) at (0.5, 0);
				\vertex (3) at (1.5, 1) {$\psi$};
				\vertex (4) at (1.5, -1) {$\overline{\psi}$};
				\vertex (h) at (0.5, 1) {$h_{\mu\nu}$};
				\vertex () at (0, -1.5) {(ii-a)};
				
				\propag [inflaton] (i) to (L);
				\propag [inflaton] (L) to (R);
				
				\propag [graviton] (L) to (h);
				
				\propag [fermion] (R) to (3);
				\propag [fermion] (4) to (R);
			\end{feynhand}
		\end{tikzpicture}
		\begin{tikzpicture}[scale=0.95]
			\begin{feynhand}
				\vertex (i) at (-1.5, 0) {$\phi$};
				\vertex (M) at (0, 0);
				\vertex (3) at (1.5, 1) {$\psi$};
				\vertex (4) at (1.5, -1) {$\overline{\psi}$};
				\vertex (h) at (1.5, 0) {$h_{\mu\nu}$};
				\vertex () at (0, -1.5) {(ii-b)};
				
				\propag [inflaton] (i) to (M);
				\propag [graviton] (0.5,0.333) to (h);            
				\propag [fermion] (M) to (3);
				\propag [fermion] (4) to (M);
			\end{feynhand}
		\end{tikzpicture}
		\begin{tikzpicture}[scale=0.95]
			\begin{feynhand}
				\vertex (i) at (-1.5, 0) {$\phi$};
				\vertex (M) at (0, 0);
				\vertex (3) at (1.5, 1) {$\psi$};
				\vertex (4) at (1.5, -1) {$\overline{\psi}$};
				\vertex (h) at (1.5, 0) {$h_{\mu\nu}$};
				\vertex () at (0, -1.5) {(ii-c)};
				\propag [inflaton] (i) to (M);
				\propag [graviton] (0.5,-0.333) to (h);            
				\propag [fermion] (M) to (3);
				\propag [fermion] (4) to (M);
			\end{feynhand}
		\end{tikzpicture}
		\begin{tikzpicture}[scale=0.95]
			\begin{feynhand}
				\vertex (i) at (-1.5, 0) {$\phi$};
				\vertex (M) at (0, 0);
				\vertex (3) at (1.5, 1) {$\psi$};
				\vertex (4) at (1.5, -1) {$\overline{\psi}$};
				\vertex (h) at (1.5, 0) {$h_{\mu\nu}$};
				\vertex () at (0, -1.5) {(ii-d)};
				\propag [inflaton] (i) to (M);
				\propag [graviton] (M) to (h);            
				\propag [fermion] (M) to (3);
				\propag [fermion] (4) to (M);
			\end{feynhand}
		\end{tikzpicture}
		
		\begin{tikzpicture}[scale=0.95]
			\begin{feynhand}
				\vertex (1) at (-1.5, 1) {$\phi$};
				\vertex (2) at (-1.5, -1) {$\psi$};
				\vertex (u) at (0, 1);
				\vertex (d) at (0, -1);
				\vertex (3) at (1.5, 1) {$h_{\mu\nu}$};
				\vertex (4) at (1.5, -1) {$\psi$};
				\vertex () at (0, -1.5) {(iii-a)};
				\propag [inflaton] (1) to (u);
				\propag [fermion] (2) to (d);
				\propag [inflaton] (u) to [edge label'=$\phi$] (d);
				\propag [graviton] (u) to (3);
				\propag [fermion] (d) to (4);
			\end{feynhand}
		\end{tikzpicture}
		\begin{tikzpicture}[scale=0.95]
			\begin{feynhand}
				\vertex (1) at (-1.5, 1) {$\phi$};
				\vertex (2) at (-1.5, -1) {$\psi$};
				\vertex (L) at (-0.5, 0);
				\vertex (R) at (0.5, 0);
				\vertex (3) at (1.5, 1) {$h_{\mu\nu}$};
				\vertex (4) at (1.5, -1) {$\psi$};
				\vertex () at (0, -1.5) {(iii-b)};
				\propag [inflaton] (1) to (L);
				\propag [fermion] (2) to (L);
				\propag [fermion] (L) to (R);
				\propag [graviton] (R) to (3);
				\propag [fermion] (R) to (4);
			\end{feynhand}
		\end{tikzpicture}
		\begin{tikzpicture}[scale=0.95]
			\begin{feynhand}
				\vertex (1) at (-1.5, 1) {$\phi$};
				\vertex (2) at (-1.5, -1) {$\psi$};
				\vertex (u) at (0, 1);
				\vertex (d) at (0, -1);
				\vertex (3) at (1.5, -1) {$h_{\mu\nu}$};
				\vertex (4) at (1.5, 1) {$\psi$};
				\vertex () at (0, -1.5) {(iii-c)};
				\propag [inflaton] (1) to (u);
				\propag [fermion] (2) to (d);
				\propag [fermion] (d) to (u);
				\propag [graviton] (d) to (3);
				\propag [fermion] (u) to (4);
			\end{feynhand}
		\end{tikzpicture}
		\begin{tikzpicture}[scale=0.95]
			\begin{feynhand}
				\vertex (1) at (-1.5, 1) {$\phi$};
				\vertex (2) at (-1.5, -1) {$\psi$};
				\vertex (M) at (0, 0);
				\vertex (3) at (1.5, 1) {$h_{\mu\nu}$};
				\vertex (4) at (1.5, -1) {$\psi$};
				\vertex () at (0, -1.5) {(iii-d)};
				\propag [inflaton] (1) to (M);
				\propag [fermion] (2) to (M);
				\propag [graviton] (M) to (3);
				\propag [fermion] (M) to (4);
			\end{feynhand}
		\end{tikzpicture}
		
		\begin{tikzpicture}[scale=0.95]
			\begin{feynhand}
				\vertex (1) at (-1.5, 1) {$\psi$};
				\vertex (2) at (-1.5, -1) {$\overline{\psi}$};
				\vertex (u) at (0, 1);
				\vertex (d) at (0, -1);
				\vertex (3) at (1.5, 1) {$h_{\mu\nu}$};
				\vertex (4) at (1.5, -1) {$A$};
				\vertex () at (0, -1.5) {(iv-a)};
				\propag [fermion] (1) to (u);
				\propag [fermion] (u) to (d);
				\propag [fermion] (d) to (2);
				\propag [photon] (d) to (4);
				\propag [graviton] (u) to (3);
			\end{feynhand}
		\end{tikzpicture}
		\begin{tikzpicture}[scale=0.95]
			\begin{feynhand}
				\vertex (1) at (-1.5, 1) {$\psi$};
				\vertex (2) at (-1.5, -1) {$\overline{\psi}$};
				\vertex (u) at (0, 1);
				\vertex (d) at (0, -1);
				\vertex (3) at (1.5, 1) {$h_{\mu\nu}$};
				\vertex (4) at (1.5, -1) {$A$};
				\vertex () at (0, -1.5) {(iv-b)};
				\propag [fermion] (1) to (u);
				\propag [fermion] (u) to (d);
				\propag [fermion] (d) to (2);
				\propag [photon] (u) to (4);
				\propag [graviton] (d) to (3);
			\end{feynhand}
		\end{tikzpicture}
		\begin{tikzpicture}[scale=0.95]
			\begin{feynhand}
				\vertex (1) at (-1.5, 1) {$\psi$};
				\vertex (2) at (-1.5, -1) {$\overline{\psi}$};
				\vertex (L) at (-0.5, 0);
				\vertex (R) at (0.5, 0);
				\vertex (3) at (1.5, 1) {$h_{\mu\nu}$};
				\vertex (4) at (1.5, -1) {$A$};
				\vertex () at (0, -1.5) {(iv-c)};
				\propag [fermion] (1) to (L);
				\propag [fermion] (L) to (2);
				\propag [photon] (L) to (R);
				\propag [graviton] (R) to (3);
				\propag [photon] (R) to (4);
			\end{feynhand}
		\end{tikzpicture}
		\begin{tikzpicture}[scale=0.95]
			\begin{feynhand}
				\vertex (1) at (-1.5, 1) {$\psi$};
				\vertex (2) at (-1.5, -1) {$\overline{\psi}$};
				\vertex (M) at (0, 0);
				\vertex (3) at (1.5, 1) {$h_{\mu\nu}$};
				\vertex (4) at (1.5, -1) {$A$};
				\vertex () at (0, -1.5) {(iv-d)};
				\propag [fermion] (1) to (M);
				\propag [fermion] (M) to (2);
				\propag [graviton] (M) to (3);
				\propag [photon] (M) to (4);
			\end{feynhand}
		\end{tikzpicture}

		\begin{tikzpicture}[scale=0.95]
			\begin{feynhand}
				\vertex (1) at (-1.5, 1) {$\psi$};
				\vertex (2) at (-1.5, -1) {$A$};
				\vertex (u) at (0, 1);
				\vertex (d) at (0, -1);
				\vertex (3) at (1.5, 1) {$h_{\mu\nu}$};
				\vertex (4) at (1.5, -1) {$\psi$};
				\vertex () at (0, -1.5) {(v-a)};
				\propag [fermion] (1) to (u);
				\propag [fermion] (u) to (d);
				\propag [photon] (d) to (2);
				\propag [fermion] (d) to (4);
				\propag [graviton] (u) to (3);
			\end{feynhand}
		\end{tikzpicture}
		\begin{tikzpicture}[scale=0.95]
			\begin{feynhand}
				\vertex (1) at (-1.5, 1) {$\psi$};
				\vertex (2) at (-1.5, -1) {$A$};
				\vertex (u) at (0, 1);
				\vertex (d) at (0, -1);
				\vertex (3) at (1.5, 1) {$h_{\mu\nu}$};
				\vertex (4) at (1.5, -1) {$\psi$};
				\vertex () at (0, -1.5) {(v-b)};
				\propag [fermion] (1) to (u);
				\propag [photon] (u) to (d);
				\propag [photon] (d) to (2);
				\propag [fermion] (u) to (4);
				\propag [graviton] (d) to (3);
			\end{feynhand}
		\end{tikzpicture}
		\begin{tikzpicture}[scale=0.95]
			\begin{feynhand}
				\vertex (1) at (-1.5, 1) {$\psi$};
				\vertex (2) at (-1.5, -1) {$A$};
				\vertex (L) at (-0.5, 0);
				\vertex (R) at (0.5, 0);
				\vertex (3) at (1.5, 1) {$h_{\mu\nu}$};
				\vertex (4) at (1.5, -1) {$\psi$};
				\vertex () at (0, -1.5) {(v-c)};
				\propag [fermion] (1) to (L);
				\propag [photon] (L) to (2);
				\propag [fermion] (L) to (R);
				\propag [graviton] (R) to (3);
				\propag [fermion] (R) to (4);
			\end{feynhand}
		\end{tikzpicture}
		\begin{tikzpicture}[scale=0.95]
			\begin{feynhand}
				\vertex (1) at (-1.5, 1) {$\psi$};
				\vertex (2) at (-1.5, -1) {$A$};
				\vertex (M) at (0, 0);
				\vertex (3) at (1.5, 1) {$h_{\mu\nu}$};
				\vertex (4) at (1.5, -1) {$\psi$};
				\vertex () at (0, -1.5) {(v-d)};
				\propag [fermion] (1) to (M);
				\propag [photon] (M) to (2);
				\propag [graviton] (M) to (3);
				\propag [fermion] (M) to (4);
			\end{feynhand}
		\end{tikzpicture}
		
		\caption{Feynman diagrams of graviton production processes considered in this
			work.\label{fig:feyn}}
	\end{figure}

	A pair of inflatons can annihilate into a pair of gravitons according
	to diagrams (i-a)-(i-d) presented in Fig.~\ref{fig:feyn}. 
	
	In this work, the inflatons are at rest so the kinematics is simple.
	Denoting the momenta of the two initial states by $p_{1}$ and $p_{2}$,
	and the final ones by $k_{1}$ and $k_{2}$, we have 
	\begin{equation}
		p_{1}^{\mu}=p_{2}^{\mu}=(m_{\phi},\vec{0})\thinspace,\ \ k_{1}^{\mu}=(m_{\phi},\vec{k})\thinspace,\ k_{2}^{\mu}=(m_{\phi},-\vec{k})\thinspace,\label{eq:A-4}
	\end{equation}
	with $\left|\vec{k}\right|=m_{\phi}$. 
	
	In general, the amplitude of each diagram can be written as
	\[
	{\cal M}=\epsilon_{\mu\nu}(k_{1})\epsilon_{\alpha\beta}(k_{2}){\cal A}^{\mu\nu,\alpha\beta}\thinspace,
	\]
	where ${\cal A}^{\mu\nu,\alpha\beta}$ denotes the remaining part
	of amplitude after the two graviton legs are removed. In practice,
	one can drop many terms in ${\cal A}^{\mu\nu,\alpha\beta}$ and use
	the following replacement:
	\begin{equation}
		\overline{{\cal A}}^{\mu\nu,\alpha\beta}\equiv\left.{\cal A}^{\mu\nu,\alpha\beta}\right|_{X\to0}\ \ \text{for}\ X\in\{\eta_{\mu\nu},\ \eta_{\alpha\beta},\ k_{1}^{\mu},\ k_{1}^{\nu},\ k_{2}^{\alpha},\ k_{2}^{\beta}\}\thinspace.\label{eq:A-5}
	\end{equation}
	Here terms proportional to $\eta_{\mu\nu}$ or $\eta_{\alpha\beta}$
	can be safely set to zero due to the traceless condition in Eq.~\eqref{eq:A}.
	Terms proportional to those $k$'s vanish due to the transverse condition
	in Eq.~\eqref{eq:A}.
	
	Using Eq.~\eqref{eq:A-5} can greatly simplify the calculation. For
	instance, ${\cal A}^{\mu\nu,\alpha\beta}$ for diagram (a) reduces
	to to a single term proportional to $p_{1}^{\mu}p_{1}^{\nu}p_{2}^{\alpha}p_{2}^{\beta}$,
	which in the nonrelativistic limit leads to $|{\cal M}_{(\text{i-a})}|^{2}=0$
	according to Eq.~\eqref{eq:A-3}. Similarly, diagram (i-b) also yields
	a null result. 
	
	For diagrams (i-c) and (i-d), a similar reduction of the amplitudes
	leads to
	\begin{align}
		\overline{{\cal A}}_{(\text{i-c})}^{\mu\nu,\alpha\beta} & =-\frac{3}{2}\frac{m_{\phi}^{2}}{M_{P}^{2}}\left(\eta^{\alpha\nu}\eta^{\beta\mu}+\eta^{\alpha\mu}\eta^{\beta\nu}\right)\thinspace,\label{eq:A-7}\\
		\overline{{\cal A}}_{(\text{i-d})}^{\mu\nu,\alpha\beta} & =2\frac{m_{\phi}^{2}}{M_{P}^{2}}\left(\eta^{\alpha\nu}\eta^{\beta\mu}+\eta^{\alpha\mu}\eta^{\beta\nu}\right)\thinspace.\label{eq:A-8}
	\end{align}
	
	From Eqs.~\eqref{eq:A-7} and \eqref{eq:A-8}, it is straightforward
	to compute the squared amplitude:
	\begin{align}
		|\mathcal{M}_{\phi\phi\rightarrow hh}|^{2} & =\left(\overline{{\cal A}}_{(\text{i-c})}^{\mu\nu,\alpha\beta}+\overline{{\cal A}}_{(\text{i-d})}^{\mu\nu,\alpha\beta}\right)\left(\overline{{\cal A}}_{(\text{i-c})}^{\mu'\nu',\alpha'\beta'}+\overline{{\cal A}}_{(\text{i-d})}^{\mu'\nu',\alpha'\beta'}\right)^{*}\mathrm{GPS}_{\mu\nu,\mu'\nu'}\mathrm{GPS}_{\alpha\beta,\alpha'\beta'}\nonumber \\
		& =\frac{2m_{\phi}^{4}}{M_{P}^{4}}\thinspace.\label{eq:A-6}
	\end{align} 
	Note that in the above calculation, we have used gravitational polarization
	summation instead of taking the average. For the latter, one should
	further multiply the result by a factor of $1/4$:
	\begin{equation}
		|\overline{\mathcal{M}}_{\phi\phi\rightarrow hh}|^{2}=|\mathcal{M}_{\phi\phi\rightarrow hh}|^{2}\times\frac{1}{4}\thinspace.\label{eq:M-div-4}
	\end{equation}
	The above
	result agrees with Eq.~(10) in Ref.~\cite{Choi:2024ilx} and Eq.~(A.18)
	in Ref.~\cite{Bernal:2025lxp}.

	\subsection{$\phi\rightarrow\psi\overline{\psi}h$}
	
	Next, we consider the gravitational bremsstrahlung from the inflaton
	decay\,---\,see diagrams (ii-a)-(ii-d) in Fig.~\ref{fig:feyn}. 
	
	We label the particle momenta as $\phi(l^{\mu})\to\psi(p^{\mu})\overline{\psi}(q^{\mu})h(k^{\mu})$
	with $l^{\mu}=(m_{\phi},\ \vec{0})$, $p^{\mu}=(E_{p},\ \vec{p})$,
	$k^{\mu}=(\omega,\ \vec{k})$, and $q^{\mu}=(m_{\phi}-E_{p}-\omega,\ -\vec{p}-\vec{k})$.
	For later use, we shall mention that 
	\begin{equation}
		p\cdot q=\frac{1}{2}m_{\phi}(m_{\phi}-2\omega)\thinspace,\label{eq:A-22}
	\end{equation}
	which is independent of $E_{p}$ for fixed $\omega$. Eq.~\eqref{eq:A-22}
	can be obtained from $q^{2}=0$ which implies $\vec{p}\cdot\vec{k}=\frac{m_{\phi}^{2}}{2}+E_{p}\omega-E_{p}m_{\phi}-m_{\phi}\omega$
	
	Using the same reduction technique in Eq.~\eqref{eq:A-5},  it is
	straightforward to simplify the matrix elements of diagrams (ii-a)-(ii-d)
	to
	\begin{align}
		i\mathcal{M}_{(\text{ii-a})} & =\,\frac{-i\,y}{2l\cdot kM_{P}}\,\left(2l_{\mu}\,l_{\nu}\right)\,\bar{u}(p)v(q)\,\epsilon^{\mu\nu}\,,\label{eq:A-9}\\
		i\mathcal{M}_{(\text{ii-b})} & =\frac{i\,y}{2p\cdot kM_{P}}\left[\bar{u}(p)(p_{\mu}\gamma_{\nu})(l\cdot\gamma)v(q)\right]\epsilon^{\mu\nu}\thinspace,\label{eq:A-10}\\
		i\mathcal{M}_{(\text{ii-c})} & =\frac{i\,y}{2q\cdot kM_{P}}\left[\bar{u}(p)(l\cdot\gamma)(q_{\mu}\gamma_{\nu})v(q)\right]\epsilon^{\mu\nu}\thinspace,\label{eq:A-11}\\
		i\mathcal{M}_{(\text{ii-d})} & \propto\eta_{\mu\nu}\epsilon^{\mu\nu}=0\,.\label{eq:A-12}
	\end{align}
	where $\mathcal{M}_{(\text{ii-d})}$ vanishes due to the traceless
	condition in Eq.~\eqref{eq:A}. In addition, according to Eq.~\eqref{eq:A-3},
	we have $|\mathcal{M}_{(\text{ii-a})}|^{2}=0$, which implies that
	$\mathcal{M}_{(\text{ii-a})}$ can be neglected. Therefore, the combined
	matrix element reads
	\begin{align}
		|\mathcal{M}_{\phi\rightarrow\psi\overline{\psi}h}|^{2} & \equiv\left|\mathcal{M}_{(\text{ii-b})}+\mathcal{M}_{(\text{ii-c})}\right|^{2}\nonumber \\
		& =\frac{y^{2}\,m_{\phi}^{2}}{M_{P}^{2}}\left(1-\frac{2\omega}{m_{\phi}}\right)\left[2-\frac{2m_{\phi}}{\omega}+\left(\frac{m_{\phi}}{\omega}\right)^{2}\right],\label{eq:A-13}
	\end{align}
	where $\omega$ is the energy of the graviton. Eq.~\eqref{eq:A-13}
	agrees with Appendix B.2 in Ref.~\cite{Barman:2023ymn} in the the
	massless fermion limit. 
	
	Similar to Eq.~\eqref{eq:M-div-4}, one should divide the above result
	by a factor of $8$ to obtain the spin/polarization-averaged matrix
	element
	\begin{equation}
		|\overline{\mathcal{M}}_{\phi\rightarrow\psi\overline{\psi}h}|^{2}=\frac{1}{8}|\mathcal{M}_{\phi\to\psi\overline{\psi}h}|^{2}\thinspace.\label{eq:M-div-8}
	\end{equation}
	
	\subsection{$\phi\psi\rightarrow\psi h$}
	
	It is also possible to produce gravitons via $\phi$-$\psi$ scattering \cite{Xu:2024fjl}.
	The relevant Feynman diagrams are shown in Fig.~\ref{fig:feyn}.
	We label the particle momenta as $\phi(l)\psi(q)\to\psi(p)h(k)$,
	and define the Mandelstam variables as $s=(l+q)^{2}$, $t=(l-p)^{2}$,
	and $u=(l-k)^{2}$. The matrix elements of the diagrams (iii-a)-(iii-d)
	are given by: 
	\begin{align}
		i\mathcal{M}_{\text{(iii-a)}} & =\,\frac{-i\,y}{2l\cdot kM_{P}}\,\left(2l_{\mu}\,l_{\nu}\right)\,\bar{u}(p)u(q)\,\epsilon^{\mu\nu}\,,\label{eq:A-14}\\
		i\mathcal{M}_{\text{(iii-b)}} & =\frac{i\,y}{2p\cdot kM_{P}}\left[\bar{u}(p)(p_{\mu}\gamma_{\nu})(l\cdot\gamma)u(q)\right]\epsilon^{\mu\nu}\thinspace,\label{eq:A-15}\\
		i\mathcal{M}_{\text{(iii-c)}} & =\frac{i\,y}{2q\cdot kM_{P}}\left[\bar{u}(p)(l\cdot\gamma)(q_{\mu}\gamma_{\nu})u(q)\right]\epsilon^{\mu\nu}\thinspace,\label{eq:A-16}\\
		i\mathcal{M}_{\text{(iii-d)}} & \propto\eta_{\mu\nu}\epsilon^{\mu\nu}=0\,.\label{eq:A-17}
	\end{align}
	We note that crossing symmetry allows us to relate these $2\to2$
	scattering matrix elements to those in the $1\to3$ process calculated
	above. Therefore, we expect that the result can be obtained from the
	previous one via crossing symmetry\footnote{See Eq.~(5.67) in Ref.~\cite{Peskin:1995ev} and also Refs.~\cite{crossing-symmetry,Bellazzini:2016xrt}.}:
	\begin{equation}
		\left|\overline{\mathcal{M}}_{\phi\psi\to\psi h}(l,p,q,k)\right|^{2}=(-1)^{\#\text{FC}}\left|\overline{\mathcal{M}}_{\phi\to\psi\overline{\psi}h}(l,p,-q,k)\right|^{2}\thinspace,\label{eq:A-19}
	\end{equation}
	where the bar over $\mathcal{M}$ indicates averaging all spins and polarizations
	in both initial and final states; the minus sign before $q$ takes
	effect on both its energy and momentum; and ``$\#\text{FC}$'' denotes
	the fermion crossing number (i.e., the number of fermions moved from
	initial to final or final to initial states)~\cite{crossing-symmetry}.
	For $\phi\psi\to\psi h$ and $\phi\to\psi\overline{\psi}h$, we have
	$\#\text{FC}=1$ and hence an overall minus sign on the right-hand
	side. However, this does not imply that $\left|\overline{\mathcal{M}}_{\phi\psi\to\psi h}(l,p,q,k)\right|^{2}$
	would be negative, since the analytic continuation of $|\overline{\mathcal{M}}_{\phi\to\psi\overline{\psi}h}|^{2}$
	from $q$ to $-q$ also introduces an extra minus sign.

	A straightforward calculation of the matrix element leads
	to 
	\begin{align}
		|\overline{\mathcal{M}}_{\phi \psi \to \psi h}|^2 =\frac{1}{8}\frac{y^2 (-2u) (m_\phi^4 + u^2) }{M_P^2 (m_\phi^2 -u)^2}\,,\label{eq:A-20}
	\end{align}
	where we have used $s+t+u=m_{\phi}^{2}$. The factor of $\frac{1}{8}$ accounts for spin/polarization-averaging of all initial and final states---similar to Eq.~\eqref{eq:M-div-8}.  
	In the non-relativistic limit of inflaton, $l = (m_\phi, \vec{0})$, $u = m_\phi^2 -2\, m_\phi\, \omega$, 
	Eq.~\eqref{eq:A-20} becomes
	\begin{align}
		|\overline{\mathcal{M}}_{\phi\psi\to\psi h}|^{2} & =\frac{1}{8}\frac{y^{2}\,m_{\phi}^{2}}{M_{P}^{2}}\left(\frac{2\omega}{m_{\phi}}-1\right)\left[2-\frac{2m_{\phi}}{\omega}+\left(\frac{m_{\phi}}{\omega}\right)^{2}\right]\,.\label{eq:A-21}
	\end{align}
	
	As one can see,  Eq.~\eqref{eq:A-21} explicitly verifies the identity of crossing symmetry in Eq.~\eqref{eq:A-19}. 
	Compared to Eqs.~\eqref{eq:M-div-8} and \eqref{eq:A-13}, Eq.~\eqref{eq:A-21} is almost the same, 
	except for an overall minus sign, which has been included in Eq.~\eqref{eq:A-19}.
	
	\subsection{$\psi\overline{\psi}\rightarrow hA$}
	
	We compute this process first in the center-of-mass (CM) frame and
	then generalize it to a general frame. We label the particle momenta
	as $\psi(p_{1})\overline{\psi}(p_{2})\to h(k_{1})A(k_{2})$ and write
	the momenta as follows:
	\begin{align}
		& p_{1}=(E,\vec{p})\thinspace,\ \ p_{2}=(E,-\vec{p})\thinspace,\label{eq:A-24}\\
		& k_{1}=(E,\vec{k})\thinspace,\ \ k_{2}=(E,-\vec{k})\thinspace,\label{eq:A-25}
	\end{align}
	with $|\vec{p}|=|\vec{k}|=E$. The Mandelstam variables therefore
	read:
	\begin{align}
		s & =2p_{1}\cdot p_{2}=2k_{1}\cdot k_{2}=4E^{2}\thinspace,\label{eq:A-26}\\
		t & =-2p_{1}\cdot k_{1}=-2p_{2}\cdot k_{2}=-2E^{2}(1-\cos\theta)\thinspace,\label{eq:A-27}\\
		u & =-2p_{2}\cdot k_{1}=-2p_{1}\cdot k_{2}=-2E^{2}(1+\cos\theta)\thinspace,\label{eq:A-28}
	\end{align}
	where $\theta$ is the angle between $\vec{p}$ and $\vec{k}$. Note
	that there are two independent kinematic parameters ($E$ and $\cos\theta$).
	Meanwhile, only two of the three Mandelstam variables are independent
	($s+t+u=0$ for this process). So the generalization from the CM frame
	to a general one can be done by mapping $E$ and $\cos\theta$ to
	two of the Mandelstam variables. 
	
	The Feynman diagrams responsible for this process are given by diagrams
	(iv-a)-(iv-d). Applying similar reduction techniques, we find that
	the amplitudes of these diagrams reduce to
	\begin{align}
		i{\cal M}_{(\text{iv-a})}= & i\kappa g\frac{\bar{v}(p_{2})\cdot\gamma^{\rho}\cdot\left[\not k_{1}\cdot\left(\gamma^{\nu}p_{1}^{\mu}+\gamma^{\mu}p_{1}^{\nu}\right)-4p_{1}^{\mu}p_{1}^{\nu}\right]\cdot u(p_{1})}{4t}\epsilon_{\mu\nu}\epsilon_{\rho}\thinspace,\label{eq:A-29}\\
		i{\cal M}_{(\text{iv-b})}= & -i\kappa g\frac{\bar{v}(p_{2})\cdot\left[\left(\gamma^{\nu}p_{2}^{\mu}+\gamma^{\mu}p_{2}^{\nu}\right)\cdot\not k_{1}-4p_{2}^{\mu}p_{2}^{\nu}\right]\cdot\gamma^{\rho}\cdot u(p_{1})}{4u}\epsilon_{\mu\nu}\epsilon_{\rho}\thinspace,\label{eq:A-30}\\
		i{\cal M}_{(\text{iv-c})}= & -\frac{g}{4s}i\kappa\ \bar{v}(p_{2})\cdot\left\lbrace 2\not k_{1}\cdot\left(\eta^{\rho\nu}\left(p_{1}^{\mu}+p_{2}^{\mu}\right)+\eta^{\rho\mu}\left(p_{1}^{\nu}+p_{2}^{\nu}\right)\right)\right.\nonumber \\
		& +\gamma^{\nu}\left(s\eta^{\rho\mu}+2k_{1}^{\mu}p_{1}^{\rho}-2k_{1}^{\rho}p_{1}^{\mu}+2k_{1}^{\mu}p_{2}^{\rho}-2k_{1}^{\rho}p_{2}^{\mu}\right)\nonumber \\
		& +\gamma^{\mu}\left(s\eta^{\rho\nu}+2k_{1}^{\nu}p_{1}^{\rho}-2k_{1}^{\rho}p_{1}^{\nu}+2k_{1}^{\nu}p_{2}^{\rho}-2k_{1}^{\rho}p_{2}^{\nu}\right)\nonumber \\
		& \left.-2\gamma^{\rho}\left[k_{1}^{\mu}\left(p_{1}^{\nu}+p_{2}^{\nu}\right)+\left(p_{1}^{\mu}+p_{2}^{\mu}\right)\left(k_{1}^{\nu}-2p_{1}^{\nu}-2p_{2}^{\nu}\right)\right]\right\rbrace \cdot u(p_{1})\epsilon_{\mu\nu}\epsilon_{\rho}\thinspace,\label{eq:A-31}\\
		i{\cal M}_{(\text{iv-d})}= & \frac{i\kappa g}{4}\bar{v}(p_{2})\cdot\left(\eta^{\rho\mu}\gamma^{\nu}+\eta^{\rho\nu}\gamma^{\mu}\right)\cdot u(p_{1})\epsilon_{\mu\nu}\epsilon_{\rho}\thinspace.\label{eq:A-32}
	\end{align}
	Averaging all the initial spins and final polarizations yields
	\begin{equation}
		|\overline{\mathcal{M}}_{\psi\overline{\psi}\rightarrow hA}|^{2}=\frac{g^{2}}{4M_{P}^{2}}E^{2}\left(3+\cos{2\theta}\right)=\frac{g^{2}}{4M_{P}^{2}}\frac{t^{2}+u^{2}}{s}\thinspace,\label{eq:M-rad-anni}
	\end{equation}
	where a factor of 1/4 accounting for spin averaging of initial fermion
	states and a factor of $1/4$ for polarization averaging of final
	states have been included. The $\frac{t^{2}+u^{2}}{s}$ part in the
	above result matches Eq.~(2.46) of Ref.~\cite{Ghiglieri:2020mhm}.
	
	\subsection{$\psi A\rightarrow h\psi$}
	
	The Feynman diagrams for this process are given in the last row of Fig.~\ref{fig:feyn}. 
	Using crossing symmetry, the squared amplitude for $\psi A\rightarrow h\psi$
	can be directly obtained from the $\psi\overline{\psi}\rightarrow hA$
	result through the Mandelstam variable substitution $s\leftrightarrow u$:
	\begin{equation}
		|\overline{\mathcal{M}}_{\psi A\rightarrow\psi h}|^{2}=\frac{g^{2}}{4M_{P}^{2}}\frac{t^{2}+s^{2}}{u}\thinspace.\label{eq:A-33}
	\end{equation}
	Note that the ordering of the particles can affect the definition
	of $s$, $t$, and $u$. We denote the momentum of $i$-th particles
	by $p_{i}$ and define $s=(p_{1}+p_{2})^{2}$, $t=(p_{2}-p_{4})^{2}$,
	$u=(p_{1}-p_{4})^{2}$. From $\psi\overline{\psi}\rightarrow hA$
	to $\psi A\rightarrow h\psi$, we perform an interchange of the 2nd
	and the 4th particle so $t$ is not changed while $s$ and $u$ are
	interchanged.
	
	One can further interchange the two initial states of $\psi A\rightarrow h\psi$
	to get the result for $A\psi\rightarrow h\psi$. Although it is essentially
	the same process, the $t$-channel enhancement in the soft-scattering
	regime (or forward scattering limit) becomes manifest:
	\begin{equation}
		|\overline{\mathcal{M}}_{A\psi\rightarrow h\psi}|^{2}=\frac{g^{2}}{4M_{P}^{2}}\frac{u^{2}+s^{2}}{t}\thinspace.\label{eq:A-34}
	\end{equation}

	\section{Calculation of collision terms\label{sec:Calc-collision}}
	
	This appendix presents a detailed calculation of the collisions terms
	used in our work. The final results are summarized as follows: 
	\begin{align}
		& {\cal C}_{h}^{\phi\phi\rightarrow hh}=\frac{\pi n_{\phi}^{2}}{32M_{P}^{4}}\delta(\omega-m_{\phi})\thinspace,\label{eq:C-1}\\
		& {\cal C}_{h}^{\phi\rightarrow\bar{\psi}\psi h}=\frac{y^{2}\rho_{\phi}}{64\pi\omega M_{P}^{2}}F\left(\frac{\omega}{m_{\phi}}\right)\Theta\left(\frac{m_{\phi}}{2}-\omega\right),\label{eq:C-2}\\
		& {\cal C}_{h}^{\phi\psi\rightarrow\psi h}=\frac{y^{2}\rho_{\phi}}{64\pi\omega M_{P}^{2}}(-1)F\left(\frac{\omega}{m_{\phi}}\right)\frac{T}{\omega}e^{-\frac{\omega-m_{\phi}/2}{T}}\Theta\left(\omega-\frac{m_{\phi}}{2}\right),\label{eq:C-3}\\
		& {\cal C}_{h}^{\bar{\psi}\psi\rightarrow Ah}=\frac{g^{2}}{12\pi^{3}M_{P}^{2}}T^{3}e^{-\omega/T}\thinspace,\label{eq:C-4}\\
		& {\cal C}_{h}^{\psi A\rightarrow\psi h}=\frac{g^{2}}{(2\pi)^{3}M_{P}^{2}}G\left(\frac{\omega}{\kappa}\right)T^{3}e^{-\omega/T}\thinspace,\label{eq:C-5}
	\end{align}
	where $\kappa=\sqrt{g^{2}n_{\psi}/T}$ is the Debye-H\"uckel screening
	scale, and
	\begin{align}
		F(x) & \equiv\left(1-2x\right)\left(2-2x^{-1}+x^{-2}\right),\label{eq:C-6}\\
		G(x) & \equiv-\frac{3}{2}-\frac{1}{4x^{2}}+\left(2+\frac{1}{2x^{2}}+\frac{1}{16x^{4}}\right)\ln\left(1+4x^{2}\right)\thinspace.\label{eq:C-7}
	\end{align}
	In the following calculation, for convenience, we use subscripts ``1'',
	``2'', ``3'', and ``4'' to denote the first, second, third,
	and fourth particles in a given process. 
	
	\subsection{$\phi\phi\rightarrow hh$}
	
	After inflation, the inflaton field forms a cold, non-relativistic
	condensate, implying that its phase space distribution  can be approximated
	by a Dirac delta function: 
	\begin{equation}
		f_{\phi}(t,\mathbf{p})=(2\pi)^{3}n_{\phi}(t)\delta^{(3)}(\mathbf{p})\thinspace.\label{eq:C-8}
	\end{equation}

	The polarization-averaged matrix element of this process is given
	by Eq.~\eqref{eq:M-div-4}, which is momentum-independent and can
	be factored out of the phase space integral in Eq.~\eqref{eq:collision-def}.
	The symmetry factor in Eq.~\eqref{eq:collision-def} for this process
	is 
	\begin{equation}
		{\cal S}=\frac{1}{2!}\times\frac{1}{1!}\thinspace,\label{eq:C-10}
	\end{equation}
	where the factor $\frac{1}{2!}$ comes from two identical $\phi$
	particles in $\phi\phi\rightarrow hh$ and the factor $\frac{1}{1!}$
	comes from gravitons. Here one might naively take two identical $h$
	particles in the final state, which would lead to a factor of $\frac{1}{2!}$.
	This would be correct if the symmetry factor is used to calculate
	the collision term for $\left[\frac{\partial}{\partial t}-3H\right]n_{h}$
	instead of $\left[\frac{\partial}{\partial t}-Hk\frac{\partial}{\partial k}\right]f_{h}$.
	For the Boltzmann equation of $f_{h}$, we have to select one $h$
	from the final states and assign it to the production state in Eq.~\eqref{eq:collision-def}.
	For this $h$, its phase space is not integrated out in Eq.~\eqref{eq:collision-def}.
	After this selection, the total number of identical particles in the
	final states reduces to one, hence the contribution is $\frac{1}{1!}$.
	For further clarifications, we refer to the discussion below Eq.~(A21)
	in Ref.~\cite{Luo:2020sho}. 
	
	The overall multiplicity factor in Eq.~\eqref{eq:collision-def} is
	\begin{equation}
		g_{n+m}=1\times1\times2\thinspace.\label{eq:C-11}
	\end{equation}
	
	Putting the above pieces together, we have
	\begin{equation}
		g_{n+m}{\cal S}|\overline{\mathcal{M}}_{\phi\phi\rightarrow hh}|^{2}=2\times\frac{1}{2!}\times\frac{1}{1!}\times\frac{2m_{\phi}^{4}}{M_{P}^{4}}\times\frac{1}{4}=\frac{m_{\phi}^{4}}{2M_{P}^{4}}\thinspace.\label{eq:C-12}
	\end{equation}
	
	Substituting Eqs.~\eqref{eq:C-12} and \eqref{eq:C-8} into Eq.~\eqref{eq:collision-def},
	the phase space integral can be computed straightforwardly: 
	\begin{align}
		{\cal C}_{h}^{\phi\phi\rightarrow hh} & =\frac{1}{2\omega}\cdot\frac{m_{\phi}^{4}}{2M_{P}^{4}}\int d\Pi_{1}d\Pi_{2}d\Pi_{3}f_{1}f_{2}(2\pi)^{4}\delta^{(4)}(p_{1}^{\mu}+p_{2}^{\mu}-p_{3}^{\mu}-p_{4}^{\mu})\nonumber \\
		& =\frac{1}{2\omega}\cdot\frac{m_{\phi}^{4}}{2M_{P}^{4}}\int\frac{n_{\phi}}{2E_{1}}\frac{n_{\phi}}{2E_{2}}d^{3}\mathbf{p}_{3}\frac{2\pi}{2E_{3}}\delta^{(4)}(p_{1}^{\mu}+p_{2}^{\mu}-p_{3}^{\mu}-p_{4}^{\mu})\nonumber \\
		& =\frac{\pi n_{\phi}^{2}m_{\phi}^{2}}{16\omega M_{P}^{4}}\int d^{3}\mathbf{p}_{3}\frac{1}{E_{3}}\delta^{(3)}(\mathbf{p}_{3}+\mathbf{p}_{4})\delta(2m_{\phi}-p_{3}-p_{4})\nonumber \\
		& =\frac{\pi n_{\phi}^{2}m_{\phi}^{2}}{16\omega M_{P}^{4}}\left[\frac{1}{E_{3}}\delta(2m_{\phi}-p_{3}-p_{4})\right]_{\mathbf{p}_{3}\to-\mathbf{p}_{4}}\nonumber \\
		& =\frac{\pi n_{\phi}^{2}m_{\phi}^{2}}{16\omega M_{P}^{4}}\left[\frac{1}{\omega}\delta(2m_{\phi}-2\omega)\right]\nonumber \\
		& =\frac{\pi n_{\phi}^{2}}{32M_{P}^{4}}\delta(\omega-m_{\phi})\thinspace.\label{eq:C-14}
	\end{align}

	\subsection{$\phi\rightarrow\bar{\psi}\psi h$}
	
	The squared amplitude of this process is given by Eq.~\eqref{eq:M-div-8}.
	Using the $F$ function in Eq.~\eqref{eq:C-6}, we rewrite it into
	a more compact form:
	\begin{equation}
		|\overline{\mathcal{M}}_{\phi\rightarrow\bar{\psi}\psi h}|^{2}=\frac{y^{2}m_{\phi}^{2}}{8M_{P}^{2}}F\left(\frac{\omega}{m_{\phi}}\right).\label{eq:C-15}
	\end{equation}
	Since it depends solely on the energy carried by the radiated graviton,
	we can factor it out from the phase space integral of the collision
	term.  So the collision term reads
	\begin{align}
		{\cal C}_{h}^{\phi\rightarrow\bar{\psi}\psi h} & =\frac{{\cal S}g_{n+m}|\overline{\mathcal{M}}_{\phi\rightarrow\bar{\psi}\psi h}|^{2}}{2\omega}\int d\Pi_{1}f_{1}d\Pi_{2}d\Pi_{3}(2\pi\delta)^{4}\nonumber \\
		& =\frac{{\cal S}g_{n+m}|\overline{\mathcal{M}}_{\phi\rightarrow\bar{\psi}\psi h}|^{2}}{2\omega}\frac{n_{\phi}}{2m_{\phi}}\int d\Pi_{2}d\Pi_{3}(2\pi\delta)^{4}\thinspace,\label{eq:C-17}
	\end{align}
	with the symmetry factor and overall multiplicity factor given by
	\begin{equation}
		{\cal S}=1,\ \ \ g_{n+m}=4\thinspace.\label{eq:C-13}
	\end{equation}
	
	The remaining part of the phase space integral is worked out at follows:
	\begin{align}
		\int d\Pi_{2}d\Pi_{3}(2\pi\delta)^{4} & =\int\frac{d^{3}\mathbf{p}_{2}}{(2\pi)^{3}2E_{2}}\frac{d^{3}\mathbf{p}_{3}}{(2\pi)^{3}2E_{3}}(2\pi\delta)^{4}\nonumber \\
		& =\frac{1}{16\pi^{2}}\int\frac{\delta\left(m_{\phi}-E_{2}(\mathbf{p}_{2})-E_{3}(-\mathbf{p}_{2}-\mathbf{p}_{4})-\omega\right)}{E_{2}(\mathbf{p}_{2})E_{3}(-\mathbf{p}_{2}-\mathbf{p}_{4})}d^{3}\mathbf{p}_{2}\nonumber \\
		& =\frac{1}{16\pi^{2}}\int\frac{d^{3}\mathbf{p}_{2}}{p_{2}\left|\mathbf{p}_{2}+\mathbf{p}_{4}\right|}\delta\left(m_{\phi}-\omega-p_{2}-\left|\mathbf{p}_{2}+\mathbf{p}_{4}\right|\right)\nonumber \\
		& =\frac{2\pi}{16\pi^{2}}\int_{0}^{\infty}p_{2}^{2}dp_{2}\int_{-1}^{1}\frac{\delta\left(m_{\phi}-\omega-p_{2}-\left|\mathbf{p}_{2}+\mathbf{p}_{4}\right|\right)}{p_{2}\left|\mathbf{p}_{2}+\mathbf{p}_{4}\right|}d\cos\theta\thinspace,\label{eq:C-16}
	\end{align}
	where $\theta$ denotes the angle between $\mathbf{p}_{2}$ and $\mathbf{p}_{4}$,
	and $\left|\mathbf{p}_{2}+\mathbf{p}_{4}\right|$ in the last step
	should be interpreted as a function of $\theta$ and $p_{2}$: 
	\begin{equation}
		\left|\mathbf{p}_{2}+\mathbf{p}_{4}\right|=\sqrt{p_{2}^{2}+\omega^{2}+2p_{2}\omega\cos\theta}\thinspace.\label{eq:C-18}
	\end{equation}
	The Dirac delta function in Eq.~\eqref{eq:C-16} is integrated out
	as follows:
	\begin{equation}
		\int_{-1}^{1}\frac{\delta\left(m_{\phi}-\omega-p_{2}-\left|\mathbf{p}_{2}+\mathbf{p}_{4}\right|\right)}{\left|\mathbf{p}_{2}+\mathbf{p}_{4}\right|}d\cos\theta=\frac{1}{p_{2}\omega},\label{eq:C-19}
	\end{equation}
	provided that the equation $m_{\phi}-\omega-p_{2}-\left|\mathbf{p}_{2}+\mathbf{p}_{4}\right|=0$
	with respect to $\cos\theta$ has a solution in the allowed range
	$[-1,1]$. Solving this equation gives rise to the following solution
	\begin{equation}
		\cos\theta=\frac{m_{\phi}^{2}+2\omega p_{2}-2m_{\phi}(\omega+p_{2})}{2\omega p_{2}}\thinspace.\label{eq:C-20}
	\end{equation}
	Imposing the condition $\cos\theta\in[-1,1]$ on Eq.~\eqref{eq:C-20},
	we obtain
	\begin{equation}
		p_{2}\in\left[\frac{m_{\phi}}{2}-\omega,\frac{m_{\phi}}{2}\right]\thinspace.\label{eq:C-21}
	\end{equation}
	Assembling the above pieces together, Eq.~\eqref{eq:C-16} becomes
	
	\begin{align}
		\int d\Pi_{2}d\Pi_{3}(2\pi\delta)^{4} & =\frac{2\pi}{16\pi^{2}}\int_{m_{\phi}/2-\omega}^{m_{\phi}/2}p_{2}^{2}dp_{2}\frac{1}{p_{2}}\frac{1}{p_{2}\omega}\Theta\left(\frac{m_{\phi}}{2}-\omega\right)\nonumber \\
		& =\frac{1}{8\pi}\Theta\left(\frac{m_{\phi}}{2}-\omega\right)\thinspace.\label{eq:C-22}
	\end{align}
	
	Hence, the collision term is given by
	\begin{align}
		{\cal C}_{h}^{\phi\rightarrow\bar{\psi}\psi h} & =\frac{{\cal S}g_{n+m}|\overline{\mathcal{M}}_{\phi\rightarrow\bar{\psi}\psi h}|^{2}}{2\omega}\frac{n_{\phi}}{2m_{\phi}}\frac{1}{8\pi}\Theta\left(\frac{m_{\phi}}{2}-\omega\right)\thinspace\nonumber \\
		& =\frac{{\cal S}g_{n+m}|\overline{\mathcal{M}}_{\phi\rightarrow\bar{\psi}\psi h}|^{2}}{32\pi\omega m_{\phi}}n_{\phi}\Theta\left(\frac{m_{\phi}}{2}-\omega\right)\nonumber \\
		& =\frac{y^{2}}{8\pi\omega}\frac{m_{\phi}n_{\phi}}{8M_{P}^{2}}F\left(\frac{\omega}{m_{\phi}}\right)\Theta\left(\frac{m_{\phi}}{2}-\omega\right).\label{eq:C-23}
	\end{align}

	\subsection{$\phi\psi\rightarrow\psi h$}
	
	The squared amplitude of this process, given by Eq.~\eqref{eq:A-21},
	can also be factored out from the phase space integral of the collision
	term. This leads to
	\begin{align}
		{\cal C}_{h}^{\phi\psi\rightarrow\psi h} & =\frac{{\cal S}g_{n+m}|\overline{\mathcal{M}}_{\phi\psi\rightarrow\psi h}|^{2}}{2\omega}\int d\Pi_{1}f_{1}d\Pi_{2}f_{2}d\Pi_{3}(2\pi\delta)^{4}\nonumber \\
		& =\frac{{\cal S}g_{n+m}|\overline{\mathcal{M}}_{\phi\psi\rightarrow\psi h}|^{2}}{2\omega}\frac{n_{\phi}}{2m_{\phi}}\int d\Pi_{2}f_{2}d\Pi_{3}(2\pi\delta)^{4}\thinspace,\label{eq:C-17-1}
	\end{align}
	The next step is similar to Eq.~\eqref{eq:C-16} except that now we
	have an extra factor of $f_{2}$ and different kinematics. So Eq.~\eqref{eq:C-16}
	changes to 
	\begin{align}
		\int d\Pi_{2}f_{2}d\Pi_{3}(2\pi\delta)^{4} & =\int\frac{d^{3}\mathbf{p}_{2}}{(2\pi)^{3}2E_{2}}\frac{d^{3}\mathbf{p}_{3}}{(2\pi)^{3}2E_{3}}(2\pi\delta)^{4}\nonumber \\
		& =\frac{2\pi}{16\pi^{2}}\int_{0}^{\infty}p_{2}^{2}dp_{2}f_{2}\int_{-1}^{1}\frac{\delta\left(m_{\phi}+p_{2}-\omega-\left|\mathbf{p}_{2}-\mathbf{p}_{4}\right|\right)}{p_{2}\left|\mathbf{p}_{2}-\mathbf{p}_{4}\right|}d\cos\theta\thinspace.\label{eq:C-16-1}
	\end{align}
	Following a similar analysis to that in the previous subsection, we
	find that Eq.~\eqref{eq:C-21} changes to
	\begin{equation}
		p_{2}\in\left[\omega-\frac{M}{2},\ \infty\right],\label{eq:C-21-1}
	\end{equation}
	while the integration of $\cos\theta$ still leads to $\frac{1}{p_{2}\omega}$.
	Therefore, ~\eqref{eq:C-16-1} becomes
	\begin{align}
		\int d\Pi_{2}f_{2}d\Pi_{3}(2\pi\delta)^{4} & =\frac{2\pi}{16\pi^{2}}\int_{\omega-\frac{m_{\phi}}{2}}^{\infty}p_{2}^{2}dp_{2}f_{2}\frac{1}{p_{2}}\frac{1}{p_{2}\omega}\Theta\left(\omega-\frac{m_{\phi}}{2}\right)\nonumber \\
		& =\frac{1}{8\pi\omega}\Theta\left(\omega-\frac{m_{\phi}}{2}\right)\int_{\omega-\frac{m_{\phi}}{2}}^{\infty}dp_{2}f_{2}\thinspace.\label{eq:C-22-1}
	\end{align}
	And the collision term reduces to
	\begin{align}
		{\cal C}_{h}^{\phi\psi\rightarrow\psi h} & =\frac{{\cal S}g_{n+m}|\overline{\mathcal{M}}_{\phi\psi\rightarrow\psi h}|^{2}}{2\omega}\frac{n_{\phi}}{2m_{\phi}}\frac{1}{8\pi\omega}\Theta\left(\omega-\frac{m_{\phi}}{2}\right)\int_{\omega-\frac{m_{\phi}}{2}}^{\infty}dp_{2}f_{2}\nonumber \\
		& =\frac{{\cal S}g_{n+m}|\overline{\mathcal{M}}_{\phi\psi\rightarrow\psi h}|^{2}n_{\phi}}{32\pi m_{\phi}\omega^{2}}\Theta\left(\omega-\frac{m_{\phi}}{2}\right)\int_{\omega-\frac{m_{\phi}}{2}}^{\infty}f_{2}dp_{2}\nonumber \\
		& \approx\frac{{\cal S}g_{n+m}|\overline{\mathcal{M}}_{\phi\psi\rightarrow\psi h}|^{2}n_{\phi}}{32\pi m_{\phi}\omega^{2}}Te^{-\frac{\omega-m_{\phi}/2}{T}}\Theta\left(\omega-\frac{m_{\phi}}{2}\right),\label{eq:C-23-1}
	\end{align}
	where in the last step we have used the Boltzmann approximation: $f_{2}\approx e^{-p_{2}/T}$.

	\subsection{$\bar{\psi}\psi\rightarrow Ah$ }
	
	The collision term of $\bar{\psi}\psi\rightarrow Ah$ where all
	particles are relativistic can be written as the following form:
	\begin{equation}
		{\cal C}_{h}^{\bar{\psi}\psi\rightarrow Ah}\approx\frac{{\cal S}g_{n+m}}{2\omega}\int d\Pi_{3}f_{1}f_{2}\int d\Pi_{1}d\Pi_{2}(2\pi\delta)^{4}|\overline{{\cal M}}_{\bar{\psi}\psi\rightarrow Ah}|^{2},\label{eq:C-24}
	\end{equation}
	where $f_{1}f_{2}$ can be treated as a quantity independent of $p_{1}$
	and $p_{2}$ using the Boltzmann approximation and energy conservation:
	\begin{equation}
		f_{1}f_{2}=e^{-(p_{1}+p_{2})/T}=e^{-(p_{3}+\omega)/T}\thinspace.\label{eq:C-25}
	\end{equation}
	
	For this process, we have
	\begin{equation}
		{\cal S}=1\thinspace,\ \ g_{n+m}=g_{\psi}g_{\bar{\psi}}g_{A}=8\thinspace.\label{eq:C-26}
	\end{equation}
	
	Next, we shall work out the integral $\int d\Pi_{1}d\Pi_{2}(2\pi\delta)^{4}|\overline{{\cal M}}_{\bar{\psi}\psi\rightarrow Ah}|^{2}$,
	which is Lorentz invariant. Therefore, one can calculate it in the
	center-of-mass (CM) frame without loss of generality. 
	
	In the CM frame, the four-momentum delta function can be written as
	\begin{equation}
		\delta^{(4)}=\delta(\sqrt{s}-E_{1}-E_{2})\delta^{(3)}(\mathbf{p}_{1}+\mathbf{p}_{2})\thinspace,\label{eq:C-29}
	\end{equation}
	where the Mandelstam variable $s$ is fixed by the final-state kinematics
	(i.e., $\mathbf{p}_{3}$ and $\mathbf{p}_{4}$) instead of $\mathbf{p}_{1}$
	and $\mathbf{p}_{2}$. So $s$ can be treated as a constant in the
	phase space integral of $\mathbf{p}_{1}$ and $\mathbf{p}_{2}$. 
	
	The matrix element in the CM frame can be written as {[}see Eq.~\eqref{eq:M-rad-anni}{]}
	\begin{equation}
		|\overline{{\cal M}}_{\bar{\psi}\psi\rightarrow Ah}|_{{\rm CM}}^{2}=\frac{g^{2}}{16M_{P}^{2}}s\left(3+\cos{2\theta}\right),\label{eq:C-28}
	\end{equation}
	where $\theta$ is the angle between $\mathbf{p}_{1}$ and $\mathbf{p}_{4}$. 
	
	Therefore, in the CM frame, the integral reads:
	\begin{align}
		\int d\Pi_{1}d\Pi_{2}(2\pi\delta)^{4}|\overline{{\cal M}}_{\bar{\psi}\psi\rightarrow Ah}|^{2} & =\frac{1}{(2\pi)^{2}}\int\frac{p_{1}^{2}dp_{1}d\Omega}{4p_{1}^{2}}\delta(\sqrt{s}-2p_{1})|\overline{{\cal M}}_{\bar{\psi}\psi\rightarrow Ah}|_{{\rm CM}}^{2},\nonumber \\
		& =\frac{1}{8(2\pi)^{2}}\int d\Omega\frac{g^{2}}{16M_{P}^{2}}s\left(3+\cos{2\theta}\right)\nonumber \\
		& =\frac{g^{2}}{48\pi M_{P}^{2}}s\thinspace.\label{eq:C-27}
	\end{align}
	Although Eq.~\eqref{eq:C-27} is derived in the CM frame, due to the
	Lorentz invariance of the integral, the result remains valid in a
	general frame. 
	
	Then the full integration proceeds as: 
	\begin{align}
		{\cal C}_{h}^{\bar{\psi}\psi\rightarrow Ah} & \approx\frac{{\cal S}g_{n+m}}{2\omega}\int d\Pi_{3}f_{1}f_{2}\frac{g^{2}}{48\pi M_{P}^{2}}s\nonumber \\
		& =\frac{{\cal S}g_{n+m}}{4}\frac{g^{2}}{24\pi M_{P}^{2}}\int\frac{p_{3}^{2}dp_{3}d\cos\theta_{3}}{(2\pi)^{3}}e^{-(\omega+p_{3})/T}(1-\cos\theta_{3})\nonumber \\
		& =\frac{{\cal S}g_{n+m}g^{2}}{12(2\pi)^{3}M_{P}^{2}}T^{3}e^{-\omega/T}\thinspace,\label{eq:C-30}
	\end{align}
	where $\theta_{3}$ denotes the angle between $\mathbf{p}_{3}$ and
	$\mathbf{p}_{4}$ and we have used $s=2\omega p_{3}(1-\cos\theta_{3})$.

	\subsection{$\psi A\rightarrow\psi h$}
	
	This process exhibits kinematic similarities to $\bar{\psi}\psi\rightarrow Ah$
	scattering, but there is a $t$-channel divergence which requires
	a careful treatment in the soft scattering limit. 
	
	First, following a similar analysis to that in the previous subsection,
	we obtain 
	\begin{equation}
		{\cal C}_{h}^{\psi A\rightarrow\psi h}\approx\frac{{\cal S}g_{n+m}}{16(2\pi)^{2}\omega}\int d\Pi_{3}e^{-(\omega+p_{3})/T}\int d\Omega|\overline{{\cal M}}_{\psi A\rightarrow\psi h}|_{{\rm CM}}^{2}\thinspace,\label{eq:C-31}
	\end{equation}
	where 
	\begin{equation}
		|\overline{{\cal M}}_{\psi A\rightarrow\psi h}|_{{\rm CM}}^{2}=\frac{g^{2}s\left(5+2\cos\theta+\cos^{2}\theta\right)}{8M_{P}^{2}(1-\cos\theta)}\thinspace,\label{eq:C-32}
	\end{equation}
	with $\theta$ the angle between $\mathbf{p}_{1}$ and $\mathbf{p}_{4}$. 
	
	As is obvious, there is a divergence at $\theta=0$ when $\cos\theta$
	is integrated from $-1$ to $1$. It originates from the divergence
	of $t\to0$ in Eq.~\eqref{eq:A-34}. Such divergences are common in
	thermal production of light particles via diagrams containing light
	$t$-channel mediators. One well-studied example is the axion production
	in thermal plasma via the Primakoff process---see e.g. Refs.~\cite{Raffelt:1985nk,Wu:2024fsf},
	where this divergence is regulated by the Debye-H\"uckel screening. 
	
	For an electrically neutral particle scattering off a electrically
	charged one, the Debye-H\"uckel screening is included by $|{\cal M}|^{2}\to|{\cal M}|^{2}{\cal F}_{{\rm Debye}}^{2}$
	with ${\cal F}_{{\rm Debye}}^{2}$ the Debye-H\"uckel form factor give
	as followed~\cite{Raffelt:1985nk,Wu:2024fsf}:  
	\begin{equation}
		{\cal F}_{{\rm Debye}}^{2}=\frac{|q^{2}|}{\kappa^{2}+|q^{2}|}\thinspace,\label{eq:C-34}
	\end{equation}
	where $q$ is the momentum transfer and $\kappa=\sqrt{g^{2}n_{\psi}/T}$
	is the Debye-H\"uckel screening scale. We note that a recent calculation
	of thermal production of gravitons in the Sun also adopted the same
	Debye-H\"uckel screening to handle the divergence~\cite{Garcia-Cely:2024ujr}. 
	
	The momentum transfer $q=p_{1}-p_{3}$ is a space-like momentum ($q^{2}<0$).
	In the CM frame, we obtain 
	\begin{equation}
		q^{2}=-2p_{1}\cdot p_{3}=-2(\omega^{2}-\mathbf{p}_{1}\cdot\mathbf{p}_{3})=-2\omega^{2}(1-c_{\theta})\thinspace,\label{eq:C-33}
	\end{equation}
	where $c_{\theta}\equiv\cos\theta$.  
	
	With the Debye-H\"uckel screening effect included, the regulated angular
	integral becomes: 
	\begin{align}
		\int d\Omega|\overline{{\cal M}}_{\psi A\rightarrow\psi h}|_{{\rm CM}}^{2}{\cal F}_{{\rm Debye}}^{2} & =\frac{2\pi g^{2}s}{8M_{P}^{2}}\int_{-1}^{1}dc_{\theta}\frac{5+2c_{\theta}+c_{\theta}^{2}}{1-c_{\theta}}\frac{2(1-c_{\theta})}{\left(\kappa/\omega\right)^{2}+2(1-c_{\theta})}\nonumber \\
		& =\frac{\pi g^{2}s}{M_{P}^{2}}\left[-\frac{3}{2}-\frac{\kappa^{2}}{4\omega^{2}}+\left(2+\frac{\kappa^{2}}{2\omega^{2}}+\frac{\kappa^{4}}{16\omega^{4}}\right)\ln\left(1+\frac{4\omega^{2}}{\kappa^{2}}\right)\right]\nonumber \\
		& =\frac{\pi g^{2}s}{M_{P}^{2}}G\left(\frac{\omega}{\kappa}\right),\label{eq:C-35}
	\end{align}
	where $G$ has been defined in Eq.~\eqref{eq:C-7}. 
	
	Finally, similar to the calculation in Eq.~\eqref{eq:C-30}, we perform
	the phase space integral of $\mathbf{p}_{3}$:
	
	\begin{align}
		{\cal C}_{h}^{\psi A\rightarrow\psi h} & \simeq\frac{{\cal S}g_{n+m}g^{2}}{32(2\pi)^{3}M_{P}^{2}}G\left(\frac{\omega}{\kappa}\right)\int p_{3}^{2}e^{-(\omega+p_{3})/T}dp_{3}\int(1-\cos\theta_{3})d\cos\theta_{3}\nonumber \\
		& =\frac{g^{2}}{(2\pi)^{3}M_{P}^{2}}G\left(\frac{\omega}{\kappa}\right)T^{3}e^{-\omega/T},\label{eq:C-30-1}
	\end{align}
	where $\theta_{3}$ denotes the angle between $\mathbf{p}_{3}$ and
	$\mathbf{p}_{4}$. 
	\bibliographystyle{JHEP}
	\bibliography{biblio}
\end{document}